\documentclass[superscriptaddress,twocolumn,amsmath,amssymb,prd,floatfix,nofootinbib]{revtex4}

\usepackage{graphicx}

\usepackage{dcolumn}
\usepackage[varg]{txfonts}
\usepackage{bm}

\begin{document}

\title{Clustering of primordial black holes formed in a
  matter-dominated epoch 
}

\author{Takahiko Matsubara} \email{tmats@post.kek.jp}
\affiliation{%
  Institute of Particle and Nuclear Studies, High Energy
  Accelerator Research Organization (KEK), Oho 1-1, Tsukuba 305-0801,
  Japan}%
\affiliation{%
  The Graduate University for Advanced Studies (SOKENDAI),
  Tsukuba, Ibaraki 305-0801, Japan}%
\author{Takahiro Terada} \email{teradat@post.kek.jp}
\affiliation{%
  Institute of Particle and Nuclear Studies, High Energy
  Accelerator Research Organization (KEK), Oho 1-1, Tsukuba 305-0801,
  Japan}%
\author{Kazunori Kohri} \email{kohri@post.kek.jp}
\affiliation{%
  Institute of Particle and Nuclear Studies, High Energy
  Accelerator Research Organization (KEK), Oho 1-1, Tsukuba 305-0801,
  Japan}%
\affiliation{%
  The Graduate University for Advanced Studies (SOKENDAI),
  Tsukuba, Ibaraki 305-0801, Japan}%
\affiliation{%
  Kavli IPMU (WPI), UTIAS, The University of Tokyo, Kashiwa,
  Chiba 277-8583, Japan}%
\author{Shuichiro Yokoyama} \email{shu@kmi.nagoya-u.ac.jp}
\affiliation{%
  Kobayashi Maskawa Institute, Nagoya University, Chikusa, Aichi
  464-8602, Japan}%
\affiliation{%
  Kavli IPMU (WPI), UTIAS, The University of Tokyo, Kashiwa,
  Chiba 277-8583, Japan}%

\date{\today}

\begin{abstract}
  In the presence of the local-type primordial non-Gaussianity, it is
  known that the clustering of primordial black holes (PBHs) emerges
  even on superhorizon scales at the formation time. This effect has
  been investigated in the high-peak limit of the PBH formation in the
  radiation-dominated epoch in the literature. There is another
  possibility that the PBH formation takes place in the early
  matter-dominated epoch. In this scenario, the high-peak limit is not
  applicable because even initially small perturbations grow and can
  become a PBH. We first derive a general formula to estimate the
  clustering of PBHs with primordial non-Gaussianity without assuming
  the high-peak limit, and then apply this formula to a model of PBH
  formation in a matter-dominated epoch. Clustering is less
  significant in the case of the PBH formation in the matter-dominated
  epoch than that in the radiation-dominated epoch. Nevertheless, it
  is much larger than the Poisson shot noise in many cases. Relations
  to the constraints of the isocurvature perturbations by the cosmic
  microwave background radiation are quantitatively discussed.
\end{abstract}

\maketitle



\section{\label{sec:Intro} Introduction }

Primordial black holes (PBHs) have recently attracted much attention
\cite{Carr:2009jm}. This is mainly because of the following reasons.
First, we can fit the signals of gravitational waves
\cite{Bird:2016dcv, Clesse:2016vqa, Sasaki:2016jop,Sasaki:2018dmp},
which have been reported by LIGO and/or Virgo. For example, in
Ref.~\cite{Abbott:2016blz}, the gravitational wave emitted from the
merger events of the binaries are fitted by assuming homogeneously
distributed PBHs with masses of
$M_{\rm PBH} \sim {\cal O}(10) M_{\odot}$. Second, PBHs with
$M_{\rm PBH} \sim {\cal O}(10^{-17}) M_{\odot}$ \cite{Carr:2009jm} --
${\cal O}(10^{-11}) M_{\odot}$ \cite{Niikura:2017zjd} can explain all
the cold dark matter (CDM) components in the Universe (see, e.g.,
Refs.~\cite{Carr:2016drx,Juan1996PhRvD,Clesse2015PBH}). Third, we can
fit the Optical Gravitational Lensing Experiment (OGLE)
ultrashort-timescale microlensing events \cite{Niikura:2019kqi} by
PBHs with their masses of
$M_{\rm PBH} \sim {\cal O}(10^{-5}) M_{\odot}$. Fourth, PBHs with
$M_{\rm PBH} \sim {\cal O}(10^{3.5})$ -- ${\cal O}(10^{5}) M_{\odot}$
may become seeds for formations of supermassive black holes (SMBHs) by
assuming a subsequent sub-Eddington accretion rate on to the seed
\cite{Kawasaki:2012kn,Kohri2014SMBH,Kawasaki:2019iis}.

Concerning a possible mechanism to produce PBHs, we expect that high
peaks of curvature perturbation (or density perturbation
$\delta \gtrsim \delta_{\rm c} \simeq 0.3$--$0.4$
\cite{Carr:1975qj,Harada:2013epa}) at small scales collapsed into PBHs
in the radiation-dominated (RD) Universe. It is known that such a high
value of curvature perturbation at small scales is produced by various
models of inflation
\cite{Inomata:2017vxo,Lyth:2011kj,Kohri:2007qn,Pi17,Gao:2018pvq},
preheating after inflation \cite{Frampton:2010sw,Martin:2019nuw}, the
curvaton in the inflationary Universe
\cite{Kawasaki:2012wr,Kohri:2012yw,Bugaev:2012ai}, Q-ball formations
\cite{Cotner:2019ykd,Kawasaki:2019iis} and so forth.

Recently, the effects of non-Gaussianities have been discussed in
investigating more precise properties of the PBH formation. For
instance, Refs.~\cite{Kawasaki:2019mbl,DeLuca:2019qsy,Young:2019yug}
focused on a nonlinear relation between the density fluctuations and
the primordial curvature perturbations on superhorizon scales, and
Ref.~\cite{Yoo:2018esr} developed a formula for the PBH abundance
with taking this nonlinearity into account in the peak theory.

As another type of non-Gaussianities, the primordial non-Gaussianity
of the curvature perturbations, which would be a probe of the
inflationary mechanism, has been also considered. It was found that
the primordial non-Gaussianity has a significant impact on the PBH
abundance (see, e.g., Refs.~\cite{Atal:2019cdz,Yoo:2019pma} and
references therein). Furthermore, some types of the primordial
non-Gaussianity could affect not only the PBH abundance but also the
spatial clustering of PBHs. There have been lots of works about this
issue. In common understanding, if the primordial curvature
perturbations obey Gaussian statistics, the distribution of the formed
PBHs would be spatially uniform; that is, the distribution is
Poissonian (see, e.g.,
Refs.~\cite{Chi06,Ali18,Desjacques:2018wuu,SY19} and references
therein). On the other hand, if the probability distribution function
of the primordial curvature perturbations would have the
non-Gaussianity which can induce the coupling between the long and
short wavelength modes, the formed PBHs would spatially clustered even
on super-Hubble scales \cite{TY15,Young:2015kda,SY19}. Such clustering
of PBHs can be observed as the matter isocurvature fluctuations in the
cosmic microwave background (CMB) and the large-scale structure, if
the PBHs are a part of the CDM component \cite{TY15,Young:2015kda}. As
another observational impact of the PBH clustering, the effect on the
merger rate of the PBH binary system, which should be an important
parameter for the LIGO/Virgo gravitational wave event, recently has
been investigated
\cite{Raidal:2017mfl,Ballesteros:2018swv,Bringmann:2018mxj,Ding:2019tjk,Vaskonen:2019jpv}.

The PBH formations are frequently assumed to take place in the RD
epoch. However, in the early Universe, oscillating energies of
nonrelativistic massive scalar fields such as the inflaton field or
curvaton field of which the energy density scales as
$\rho \propto a^{-3}$ with scale factor $a=a(t)$ may dominate the
energy density of the Universe until their decays (i.e., until the
reheating time). In this case, an early matter-dominated (MD) epoch
could be realized before the RD epoch.

More concretely, moduli or dilaton fields, which are predicted in
particle physics models beyond the standard model such as supergravity
and/or superstring theory, tend to have a long lifetime. That is
because they decay only through gravitational interaction. For
example, with their masses of the order of weak scale, the lifetime
can be ${\cal O}(1)~{\rm sec}$ and reheating temperature after its
domination becomes $T_\mathrm{R} \sim O(1)$ MeV.~\cite{Hasegawa:2019jsa} (see
also
Refs.~\cite{Kawasaki:1999na,Kawasaki:2000en,Hannestad:2004px,Ichikawa:2005vw,deSalas:2015glj}).
In this case, PBHs with their masses up to
$M_{\rm PBH} \lesssim {\cal O}(10^{3})\, M_{\odot}$ could be produced
in the early MD epoch.

In this paper, we investigate the clustering property of the PBHs
formed in the early MD epoch in the presence of local-type
non-Gaussianity. In Refs.~\cite{TY15,Young:2015kda,SY19}, which focus
on the PBH formation in the RD epoch, a simple high-peak formalism is
employed to evaluate the two-point correlation function or the power
spectrum of the spatial fluctuations of PBH number density, which
characterize the PBH clustering. This is because in the RD epoch PBHs
are considered to be simply formed through the spherical gravitational
collapse of the overdense region with Hubble scales. On the other
hand, the formation of PBHs in the early MD epoch are completely
different from the ones in the RD epoch. Because perturbations evolve
nonspherically in MD epochs under negligible pressure, even if
$ \delta \ll \delta_{\rm c}$, a PBH can form once a region is enclosed
by its event horizon~\cite{Khlopov:1980mg,Polnarev:1982,Har16}. By
considering finite angular momentum in each patch of horizon, the
number density of the PBHs produced in the early MD epoch is
suppressed exponentially due to their own spins~\cite{Har17} (see also
Ref.~\cite{Kokubu:2018fxy} for an additional suppression of the number
density due to inhomogeneities). Therefore, it is a nontrivial
question which dominates between the clustering of PBHs and the
Poisson noise of the PBHs formed in the MD epoch.

We employ a model of Refs.~\cite{Har16,Har17} for the PBH formation in
the MD epoch, taking into account the nonlinear, nonspherical
evolutions of the matter density with the Zel'dovich approximation
\cite{Zel70} and the PBH formation with the hoop conjecture
\cite{Tho72}. In order to carefully treat the details of the formation
process, we make use of a method of the integrated perturbation theory
(iPT) \cite{Mat95,Mat11,Mat12,YM13,Mat14}. The iPT is a general
framework to predict the clustering properties of biased fields and is
able to take into account the effects of nonlinear evolutions of
clustering, redshift-space distortions, primordial non-Gaussianity,
etc. In this paper, we are interested in the clustering of PBHs at the
formation time, and the iPT is used only in estimating a contribution
of primordial non-Gaussianity to the initial clustering of PBHs.

The paper is organized as follows. In Sec.~\ref{sec:PS}, a general
consequence of the iPT for the initial power spectrum of PBHs due to
the primordial local-type non-Gaussianity is summarized. It is shown
that the iPT can successfully reproduce the previous results in the
high-peak limit. In Sec.~\ref{sec:MDPBH}, the formula for the PBH
clustering in the MD epoch, which is a main result of this paper, is
derived. Observational implications are discussed in
Sec.~\ref{sec:Observational}. Our conclusions are given in
Sec.~\ref{sec:Conclusions}. Technical details of the derivation of our
formulas are given in Appendixes~\ref{app:RBF} and
\ref{app:AnalyticRBF}. Detailed discussion on the observational
constraints is given in Appendix~\ref{app:constraints}.

\section{\label{sec:PS} Initial power spectrum of PBHs with primordial
  non-Gaussianity }

The PBHs are considered to be biased objects of the energy density in
the early Universe. In this section, we generally consider the biased
power spectrum in the presence of local-type non-Gaussianity, by
making use of the iPT formalism. The results of this section is valid
in both the RD and MD epochs.

\subsection{\label{subsec:LocalnG}
Local-type non-Gaussianity}

We consider the primordial non-Gaussianity characterized by
higher-order polyspectra of the curvature perturbations on the
comoving slice $\mathcal{R}$ as
\begin{align}
  \label{eq:2-1a}
&  \left\langle
  \mathcal{R}(\bm{k}_1)\mathcal{R}(\bm{k}_2)
  \right\rangle_\mathrm{c}
  = (2\pi)^3 \delta_\mathrm{D}^3(\bm{k}_1 + \bm{k}_2)
    P_\mathcal{R}(k_1),
  \\
  \label{eq:2-1b}
&  \left\langle
  \mathcal{R}(\bm{k}_1)\mathcal{R}(\bm{k}_2)\mathcal{R}(\bm{k}_3)
  \right\rangle_\mathrm{c}
  = (2\pi)^3 \delta_\mathrm{D}^3(\bm{k}_1 + \bm{k}_2 + \bm{k}_3)
    B_\mathcal{R}(\bm{k}_1,\bm{k}_2,\bm{k}_3),
  \\
&  \left\langle
     \mathcal{R}(\bm{k}_1)\mathcal{R}(\bm{k}_2)\mathcal{R}(\bm{k}_3)
     \mathcal{R}(\bm{k}_4) 
  \right\rangle_\mathrm{c}
  = (2\pi)^3 \delta_\mathrm{D}^3(\bm{k}_1+\bm{k}_2+\bm{k}_3+\bm{k}_4)
\nonumber\\
  \label{eq:2-1c}
  & \hspace{11pc}
  \times  T_\mathcal{R}(\bm{k}_1,\bm{k}_2,\bm{k}_3,\bm{k}_4),
\end{align}
where $\langle\cdots\rangle_\mathrm{c}$ denotes the cumulant, or the
connected part of correlations, and $P_\mathcal{R}$, $B_\mathcal{R}$,
and $T_\mathcal{R}$ are called the power spectrum, bispectrum, and
trispectrum, respectively. For the local-type non-Gaussianity, the
higher-order polyspectra are given by \cite{BSW06}
\begin{align}
  \label{eq:2-2a}
&  B_\mathcal{R}(\bm{k}_1,\bm{k}_2,\bm{k}_3)
  = \frac{6}{5} f_\mathrm{NL}
  \left[
    P_\mathcal{R}(k_1)P_\mathcal{R}(k_2) + \mathrm{cyc.}
    \right],
  \\
&  T_\mathcal{R}(\bm{k}_1,\bm{k}_2,\bm{k}_3,\bm{k}_4)
  = \frac{54}{25} g_\mathrm{NL}
  \left[
    P_\mathcal{R}(k_1)P_\mathcal{R}(k_2)P_\mathcal{R}(k_3)
    + \mathrm{3\ perms}
    \right]
\nonumber\\
  \label{eq:2-2b}
  & \hspace{4.5pc}
    + \tau_\mathrm{NL}
  \left[
    P_\mathcal{R}(k_1)P_\mathcal{R}(k_2)P_\mathcal{R}(k_{13})
    + \mathrm{11\ perms}
    \right],
\end{align}
where $k_{13} \equiv |\bm{k}_1+\bm{k}_3|$ etc., and $f_\mathrm{NL}$,
$g_\mathrm{NL}$, and $\tau_\mathrm{NL}$ are the parameters of
local-type non-Gaussianity, and perms stands for permutations and cyc
stands for cyclic permutations. If the primordial curvature
perturbations emerge from the quantum fluctuations of a single scalar
field, there is a relation,
$\tau_\mathrm{NL} = (36/25)
{f_\mathrm{NL}}^2$~\cite{Boubekeur:2005fj}. If multiple scalar fields
contribute, there is an inequality,
$\tau_\mathrm{NL} > (36/25) {f_\mathrm{NL}}^2$ \cite{SY08}.

The relation at linear order between comoving curvature perturbations
$\mathcal{R}$ and the linear density contrast $\delta_\mathrm{L}$ on
comoving slices is given by \cite{LL00,YBS14}
\begin{equation}
  \label{eq:2-3}
  \delta_\mathrm{L}(\bm{k}) = 
  \mathcal{M}(k) \mathcal{R}(\bm{k}),
\end{equation}
where the proportional factor in the RD and MD epochs is given by
\begin{equation}
  \label{eq:2-4}
  \mathcal{M}(k) \equiv
  \frac{2+2w}{5+3w} \frac{k^2 T(k)}{a^2H^2},
\end{equation}
where $w=p/\rho$ is the parameter of the equation of state and
$H=\dot{a}/a$ is the Hubble parameter. The transfer function $T(k)$
describes the evolution on subhorizon scales, and the time
dependencies in various functions are suppressed in our notations for
simplicity. For example,
$T(k,\eta) =[\sin(k\eta) - k\eta\cos(k\eta)]/[3(k\eta)^3]$ in the RD
epoch, where $\eta$ is the conformal time. In the applications to PBHs
in the following sections, we are interested in the superhorizon
scales at the formation epoch of PBHs, where we can safely put
$T(k) = 1$.

\subsection{\label{subsec:iPTPS} Power spectrum in the presence of
  local-type non-Gaussianity }

In the iPT formalism, the renormalized bias functions in Lagrangian
space are defined by \cite{Mat11,Mat12}
\begin{equation}
  \label{eq:2-10}
  c^\mathrm{L}_n(\bm{k}_1,\ldots,\bm{k}_n) =
  (2\pi)^{3n} \int\frac{d^3k}{(2\pi)^3}
  \left\langle
    \frac{\delta^n \delta^\mathrm{L}_X(\bm{k})}
    {\delta\delta_\mathrm{L}(\bm{k}_1)\cdots\delta\delta_\mathrm{L}(\bm{k}_n)}
  \right\rangle,  
\end{equation}
where $\delta^\mathrm{L}_X(\bm{k})$ is the density contrast of the
biased objects in Lagrangian space as a functional of the linear
density contrast, and $\delta/\delta\delta_\mathrm{L}(\bm{k})$ is a
functional derivative.

The iPT formalism applies to any biased objects in general, while in
this paper we identify the biased objects as PBHs in later sections.
In Ref.~\cite{YM13}, the power spectrum of biased objects in the
large-scale limit, where nonlinear evolution of the matter density
field is negligible, is calculated by the formalism of iPT in the
late-time MD epoch. Substituting $\mathcal{M}(k)$ of this literature
by $(5/3)\mathcal{M}(k)$ in this paper, the same expressions as
Eqs.~(23) and (34) of Ref.~\cite{YM13} hold in both the RD and MD
epochs. Thus the result of iPT for the power spectrum of the biased
objects in the large-scale limit is given by
\begin{align}
  P_X(k)
  \simeq
  &
    [b_1(k)]^2 P_\mathrm{L}(k)
    \nonumber\\
  &+
  \frac{12}{5} f_\mathrm{NL} b_1(k)
  \frac{P_\mathrm{L}(k)}{\mathcal{M}(k)}
  \int \frac{d^3p}{(2\pi)^3}
  c^\mathrm{L}_2(\bm{p},-\bm{p})P_\mathrm{L}(p)
  \\
  & +
  \left(\frac{54}{25}g_\mathrm{NL} + 2 \tau_\mathrm{NL}\right)
  b_1(k) \frac{P_\mathrm{L}(k)}{\mathcal{M}(k)}
    \int \frac{d^3p_1}{(2\pi)^3} \frac{d^3p_2}{(2\pi)^3}
    \nonumber\\
  & \quad \times
  c^\mathrm{L}_3(\bm{p}_1,\bm{p}_2,-\bm{p}_1-\bm{p}_2)
  \frac{\mathcal{M}(p_{12})}{\mathcal{M}(p_1)\mathcal{M}(p_2)}
  P_\mathrm{L}(p_1) P_\mathrm{L}(p_2)
  \nonumber\\
  & + \tau_\mathrm{NL} \frac{P_\mathrm{L}(k)}{\mathcal{M}^2(k)}
  \left[
    \int \frac{d^3p}{(2\pi)^3} c^\mathrm{L}_2(\bm{p},-\bm{p})
    P_\mathrm{L}(p)
    \right]^2
  \nonumber\\
  & + P_\mathrm{const},
  \label{eq:2-11}
\end{align}
where $P_\mathrm{L}(k)$ is the power spectrum of the linear density
field $\delta_\mathrm{L}$, $b_1(k) \equiv 1+c^\mathrm{L}_1(k)$ is the
linear bias parameter in Eulerian space, and $P_\mathrm{const}$ is the
higher-order correction terms which are constant in the large-scale
limit of $k\rightarrow 0$.

The most dominant contribution in the large-scale limit of
$k\rightarrow 0$ is given by the term with a factor
$\mathcal{M}^{-2}(k)$, because $\mathcal{M}(k) \sim k^2$. The
corresponding term of the most dominant contribution is the last term
but one in Eq.~(\ref{eq:2-11}). The factor $\mathcal{M}(k)$ is
sufficiently small for $k \ll aH =R^{-1}$, where $R$ is the comoving
horizon scale which gives the mass scale $M \sim 4\pi R^3/3$ of PBHs
in later sections. Therefore, the most dominant term of the power
spectrum in the large-scale limit $k \ll R^{-1}$ is given by
\begin{equation}
  \label{eq:2-12}
  P_X(k) \simeq
  \tau_\mathrm{NL} {C_2}^2 \frac{P_\mathrm{L}(k)}{\mathcal{M}^2(k)} =
  \tau_\mathrm{NL} {C_2}^2 P_{\mathcal{R}}(k),
\end{equation} 
where
\begin{equation}
  \label{eq:2-13}
  C_2 \equiv
  \int \frac{d^3p}{(2\pi)^3} c^\mathrm{L}_2(\bm{p},-\bm{p})
  P_\mathrm{L}(p).
\end{equation}
Equation~(\ref{eq:2-12}) is the general prediction of iPT for the
biased power spectrum with local-type non-Gaussianity in the
large-scale limit of $k\rightarrow 0$. There appears a strongly
scale-dependent bias,
$P_X(k)/P_\mathrm{L}(k) \propto \mathcal{M}^{-2}(k) \propto k^{-4}$ in
the large-scale limit. The same scaling property is also derived from
the peak-background split in the halo model \cite{SL11,BFGS13}. The
amplitude of the scale-dependent bias is proportional to the product
$\tau_\mathrm{NL} {C_2}^2$, and the factor $C_2$ depends on the
formation process of the biased objects.

\subsection{\label{subsec:highpeaks} High-peak limit of thresholded
  regions }

The result of Eq.~(\ref{eq:2-12}) in the previous subsection is quite
general for any biased objects. To determine the amplitude, the factor
$C_2$ should be estimated. This factor has a simple form in a
high-peak limit, which we first consider here. The high-peak limit of
thresholded regions is frequently considered as an approximation of
formation sites of PBHs in the RD epoch \cite{Chi06}. The number
density of the collapsed objects above a threshold $\delta_\mathrm{c}$
is given by
\begin{equation}
  \label{eq:2-20}
  n_\mathrm{th}(\bm{x})
  = \frac{\bar{n}_\mathrm{th}}{\beta^\mathrm{th}_0}
  \Theta\left[\delta_R(\bm{x})-\delta_\mathrm{c}\right],
\end{equation}
where $\Theta(x)$ is the Heaviside step function,
\begin{equation}
  \label{eq:2-22}
  \delta_R(\bm{x})
  = \int \frac{d^3k}{(2\pi)^3}
  e^{i\bm{k}\cdot\bm{x}} \delta_\mathrm{L}(\bm{k})W(kR)
\end{equation}
is the smoothed density contrast with a smoothing radius $R$, $W(kR)$
is the window function, $\bar{n}_\mathrm{th} = \langle
n_\mathrm{th}(\bm{x})\rangle$ is the mean number density, and 
\begin{equation}
  \label{eq:2-23}
  \beta^\mathrm{th}_0 \equiv
  \left\langle
    \Theta\left(\delta_R-\delta_\mathrm{c}\right)
  \right\rangle
\end{equation}
is the production probability. The number density of
Eq.~(\ref{eq:2-20}) is an example of the local Lagrangian bias, and
the renormalized bias in this case is given by (Eq.~(89) of
Ref.~\cite{Mat11})
\begin{equation}
  \label{eq:2-24}
  c^\mathrm{L}_n(\bm{k}_1,\ldots,\bm{k}_n) = \frac{1}{\bar{n}_\mathrm{th}}
  \left\langle
    \frac{d^n n_\mathrm{th}}{d{\delta_R}^n}
  \right\rangle
  W(k_1R) \cdots W(k_nR).
\end{equation}
Specifically, we have
\begin{align}
  c^\mathrm{L}_n(\bm{k}_1,\ldots,\bm{k}_n) =
  \frac{\beta^\mathrm{th}_n}{\beta^\mathrm{th}_0}
  W(k_1R) \cdots W(k_nR),
  \label{eq:2-25}
\end{align}
where
\begin{equation}
  \label{eq:2-26}
  \beta^\mathrm{th}_n \equiv
  \left\langle
    \delta_\mathrm{D}^{(n-1)}\left(\delta_R-\delta_\mathrm{c}\right)
  \right\rangle,
\end{equation}
and
$\delta_\mathrm{D}^{(n-1)}(x) = d^{n-1}\delta_\mathrm{D}(x)/dx^{n-1} =
d^n\Theta(x)/dx^n$ is the $(n-1)$th derivative of the Dirac delta
function, $\delta_\mathrm{D}(x) = d\Theta(x)/dx$.

Up to the lowest order in non-Gaussianity parameters, the averages of
Eqs.~(\ref{eq:2-23}) and (\ref{eq:2-26}) can be estimated with
Gaussian statistics, provided that they are substituted in
Eqs.~(\ref{eq:2-12}) and (\ref{eq:2-13}). Using the variance of the
smoothed density contrast,
\begin{equation}
  \label{eq:2-27}
  \sigma^2 \equiv
  \int\frac{k^2dk}{2\pi^2} P_\mathrm{L}(k) W^2(kR),
\end{equation}
we have
\begin{align}
  \beta^\mathrm{th}_n
  &= \frac{1}{\sqrt{2\pi}\sigma}
  \int_{-\infty}^\infty d\delta\, e^{-\delta^2/(2\sigma^2)}
    \delta_\mathrm{D}^{(n-1)}\left(\delta-\delta_\mathrm{c}\right)
  \nonumber\\
  &= \frac{1}{\sqrt{2\pi} \sigma^n}
    H_{n-1}(\nu)
    e^{-\nu^2/2},
  \label{eq:2-28}
\end{align}
where $\nu \equiv \delta_\mathrm{c}/\sigma$, and
$H_m(x) = e^{x^2/2}(-d/dx)^n e^{-x^2/2}$ is the Hermite polynomial.
If we define
\begin{equation}
  \label{eq:2-29}
  H_{-1}(x) \equiv
  \sqrt{\frac{\pi}{2}} e^{x^2/2}
  \mathrm{erfc}\left(\frac{x}{\sqrt{2}}\right),  
\end{equation}
Eq.~(\ref{eq:2-28}) holds also in the case of $n=0$ \cite{Mat95}.
Substituting Eq.~(\ref{eq:2-25}) with $n=2$ into Eq.~(\ref{eq:2-13}),
we have
\begin{equation}
  \label{eq:2-29-1}
  C_2  = \frac{H_1(\nu)}{H_{-1}(\nu)},
\end{equation}
and Eq.~(\ref{eq:2-12}) reduces to
\begin{align}
  P_\mathrm{PBH}(k)
  &=
  \left(\frac{H_1(\nu)}{H_{-1}(\nu)}\right)^2
  \tau_\mathrm{NL}
    \frac{P_\mathrm{L}(k)}{\mathcal{M}^2(k)}
    \nonumber\\
  &=
  \frac{2\nu^2}{\pi e^{\nu^2}}
  \left[\mathrm{erfc}\left(\frac{\nu}{\sqrt{2}}\right)\right]^{-2}
  \tau_\mathrm{NL}
  \frac{P_\mathrm{L}(k)}{\mathcal{M}^2(k)},
  \label{eq:2-30}
\end{align}
when the biased objects are identified as PBHs. 

The PBH formation in the RD epoch is frequently modeled by a high-peak
limit $\nu \rightarrow \infty$ of the thresholded regions. In the
high-peak limit, we have $H_m(\nu) \rightarrow \nu^m$, including
$m=-1$. In this limit, we have $C_2 \simeq \nu^2$, and
Eq.~(\ref{eq:2-30}) reduces to a simple expression,
\begin{equation}
  \label{eq:2-31}
  P_\mathrm{PBH}(k) \simeq
  \nu^4 \tau_\mathrm{NL}\frac{P_\mathrm{L}(k)}{\mathcal{M}^2(k)}
  = \nu^4  \tau_\mathrm{NL}P_\mathcal{R}(k).
\end{equation}
This equation is consistent with the results of
Refs.~\cite{TY15,SY19}.\footnote{The definition of $f_\mathrm{NL}$ in
  Ref.~\cite{TY15} corresponds to $(3/5)f_\mathrm{NL}$ in most of
  literature and in this paper.}

\section{\label{sec:MDPBH} Initial clustering of PBHs formed in a
  matter-dominated epoch in the presence of primordial
  non-Gaussianity}

In the previous section, we found that the dominant contribution in
the large-scale limit to the initial power spectrum is given by
Eq.~(\ref{eq:2-12}) in the presence of local-type non-Gaussianity. In
that expression, the integral $C_2$ of Eq.~(\ref{eq:2-13}), together
with non-Gaussianity parameter $\tau_\mathrm{NL}$, determines the
amplitude of the initial power spectrum of PBHs. As noted in the last
section, this integral in the high-peak limit is given by
$C_2 \simeq \nu^2$, when the PBH is assumed to form with a condition,
$\delta \geq \nu\sigma$. The high-peak limit is satisfied in a usual
assumption that the PBH formed in the RD epoch where
$\nu \sim \mathcal{O}(10)$ \cite{Chi06,TY15,Ali18}. However, there is
a possibility that the high-peak limit is not satisfied in the PBH
formation. For example, there are scenarios in which the PBHs are
formed in a MD epoch \cite{Har16,Har17}, in which case the high-peak
limit is not appropriate. In the PBH formation in a MD epoch,
nonspherical effects in gravitational collapse play a crucial role.

In this section, we apply a model of Refs.~\cite{Har16,Har17}. In the
model, the Zel'dovich approximation \cite{Zel70}, Thorne's hoop
conjecture \cite{Tho72}, and Doroshkevich's probability distribution
\cite{Dor70} are combined to predict the PBH formation in a MD epoch.

\subsection{\label{subsec:Model} Model of PBH formation in a MD epoch
}

We apply a model of Ref.~\cite{Har16} for the PBH formation in a MD
epoch. In this model, the criteria of black hole formation is given by
\begin{equation}
  \label{eq:3-1}
  h(\alpha,\beta,\gamma) \leq 1, \quad \alpha>0,
\end{equation}
where
\begin{equation}
  \label{eq:3-2}
  h(\alpha,\beta,\gamma) =
  \frac{2}{\pi} \frac{\alpha-\gamma}{\alpha^2}
  E\left[\sqrt{1 - \left(\frac{\alpha-\beta}{\alpha-\gamma}\right)^2}\right],
\end{equation}
and $\alpha \geq \beta \geq \gamma$ are eigenvalues of the
inhomogeneous part of the deformation tensor in the Zel'dovich
approximation. They are eigenvalues of a tensor
$\varphi_{ij} \equiv \partial_i\partial_j\varphi$, where $\varphi$ is
a normalized linear potential, $\triangle\varphi = \delta_R$, and
$\delta_R$ is the smoothed linear density perturbations with smoothing
radius $R$, and the smoothing radius corresponds to the mass scale of
the PBH. The function $E(k)$ in Eq.~(\ref{eq:3-2}) is the complete
elliptic integral of the second kind, and is a monotonically
decreasing function of $0\leq k\leq 1$. The above criterion is derived
by combining the Zel'dovich approximation \cite{Zel70} and the hoop
conjecture \cite{Tho72} for the PBH formation in the MD epoch. See
Ref.~\cite{Har16} for the details of the derivation of the above
condition.

As mentioned in the previous section, we can assume that each
independent component of the deformation tensor in the Zel'dovich
formula obeys Gaussian statistics up to the lowest order in
non-Gaussianity parameters when the factor $C_2$ in
Eq.~(\ref{eq:2-13}) is evaluated. The probability distribution of
$\alpha$, $\beta$, and $\gamma$ is given by \cite{Dor70}
\begin{multline}
  \label{eq:3-3}
  w(\alpha,\beta,\gamma) =
  \frac{3^3 5^{5/2}}{8 \pi \sigma^6}
  (\alpha-\beta)(\beta-\gamma)(\alpha-\gamma)
  \\ \times
  \exp\left[
    -\frac{3}{\sigma^2}(\alpha^2 + \beta^2 + \gamma^2)
    +\frac{3}{2\sigma^2}(\alpha\beta + \beta\gamma + \gamma\alpha)
  \right],
\end{multline}
where $\sigma^2 = \langle {\delta_R}^2\rangle$ is the
variance of smoothed linear density perturbations.

\begin{figure}
\centering
\includegraphics[height=12.5pc]{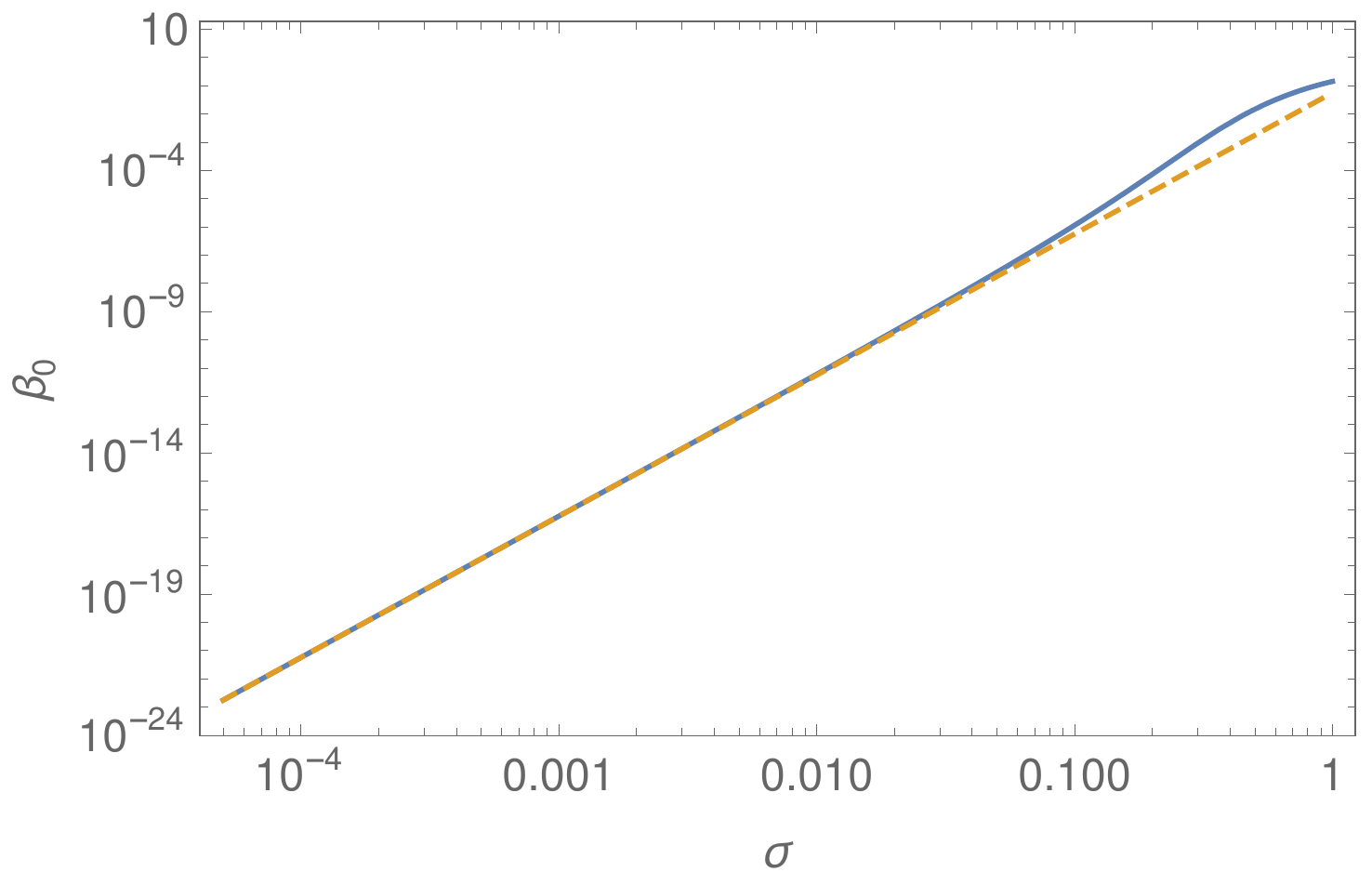}
\caption{\label{fig:beta0wo} Production probability $\beta_0$ in a
  model of Ref.~\cite{Har16}. The blue solid line corresponds to the
  result of the numerical integration of Eq.~(\ref{eq:3-5}), and the
  orange dashed line corresponds to the analytic estimate of
  Eq.~(\ref{eq:3-11}) for $\sigma \ll 1$.}
\end{figure}

According to the above criteria, the number density of PBH is given by
\begin{equation}
  \label{eq:3-4}
  n_\mathrm{PBH}(\alpha,\beta,\gamma) =
  \frac{\bar{n}_\mathrm{PBH}}{\beta_0} \Theta(\alpha)
  \Theta\left[1-h(\alpha,\beta,\gamma)\right],
\end{equation}
where $\beta_0$ is the production probability of PBH \cite{Har16},
\begin{equation}
  \label{eq:3-5}
  \beta_0 = \int_0^\infty d\alpha \int_{-\infty}^\alpha d\beta
   \int_{-\infty}^\beta d\gamma
   \Theta\left[1-h(\alpha,\beta,\gamma)\right]
   w(\alpha,\beta,\gamma).
\end{equation}
In Fig.~\ref{fig:beta0wo}, the production probability $\beta_0$ is
plotted as a function of $\sigma$. This figure reproduces the
corresponding result of Fig.~1 of Ref.~\cite{Har16}. 
In Ref.~\cite{Har16}, an analytic estimate of the function $\beta_0$
of Eq.~(\ref{eq:3-5}) is given for $\sigma \ll 1$. The result is given
by
\begin{equation}
  \label{eq:3-11}
  \beta_0
  \simeq 0.05556\sigma^5.
\end{equation}
This asymptotic formula is also plotted in Fig.~\ref{fig:beta0wo}.

\subsection{\label{subsec:cncalc}
  Calculating the renormalized bias functions
}

In this subsection, we explicitly calculate the renormalized bias
functions of orders 1 and 2. The derivation is quite similar to the
one described in Ref.~\cite{MD16}, where renormalized bias functions
of peaks are calculated. The derivations are quite technical and the
detailed calculations are given in Appendixes~\ref{app:RBF} and
\ref{app:AnalyticRBF}.

The renormalized bias functions $c_n$ in general are calculated by
a definition of Eq.~(\ref{eq:2-10}), which are equivalent to an
expression, 
\begin{equation}
  \label{eq:3-7}
  c_n(\bm{k}_1,\ldots,\bm{k}_n) =
  \frac{(2\pi)^{3n}}{\bar{n}_\mathrm{PBH}}
  \left\langle
    \frac{\delta^n n_\mathrm{PBH}[\delta_\mathrm{L}]}
    {\delta\delta_\mathrm{L}(\bm{k}_1)\cdots\delta\delta_\mathrm{L}(\bm{k}_n)}
  \right\rangle,
\end{equation}
where $n_\mathrm{PBH}[\delta_\mathrm{L}]$ is a number density of PBHs
at any point as a functional of $\delta_\mathrm{L}$.
The number density of Eq.~(\ref{eq:3-4}) is a function of a tensor
$\varphi_{ij}$, and thus is a functional of linear density field
$\delta_\mathrm{L}$. In Fourier space, relations among variables are
given by
\begin{equation}
  \label{eq:3-6}
  \delta_R(\bm{k}) = W(kR) \delta_\mathrm{L}(\bm{k}), \quad
  \varphi_{ij}(\bm{k}) = \hat{k}_i \hat{k}_j W(kR) \delta_\mathrm{L}(\bm{k}),
\end{equation}
where $\hat{\bm{k}} = \bm{k}/|\bm{k}|$.

The detailed derivation of the renormalized bias functions of $c_1$
and $c_2$ with our model of Eq.~(\ref{eq:3-4}) is given in
Appendix~\ref{app:RBF}. As a result, the renormalized bias functions
up to second order are given by
\begin{align}
  \label{eq:3-8a}
  c_1(\bm{k})
  &= b^\mathrm{L}_1 W(kR),
  \\              
  \label{eq:3-8b}
  c_2(\bm{k}_1,\bm{k}_2)
  &=
    \left\{
    b^\mathrm{L}_2 + 
    \left[
    3 (\hat{\bm{k}}_1\cdot\hat{\bm{k}}_2)^2 - 1
    \right] \omega^\mathrm{L}_1
    \right\}
    W(k_1R)W(k_2R),
\end{align}
where $b^\mathrm{L}_1$, $b^\mathrm{L}_2$, and $\omega^\mathrm{L}_1$
are given by Eqs.~(\ref{eq:a-27a})--(\ref{eq:a-27c}), with
Eqs.~(\ref{eq:a-26}) and (\ref{eq:3-5}). The integrals of
Eqs.~(\ref{eq:a-27a})--(\ref{eq:a-27c}) and (\ref{eq:3-5}) can be
numerically evaluated in general. The necessary numerical integrations
reduce to virtually two-dimensional ones by transformations which are
described in Appendix~\ref{app:AnalyticRBF}, and explicitly given by
Eq.~(\ref{eq:b-6}) with Eqs.~(\ref{eq:b-5}) and (\ref{eq:b-7}).

Using the similar technique of Ref.~\cite{Har16}, one can obtain
analytic estimates for $b^\mathrm{L}_1$, $b^\mathrm{L}_2$, and
$\omega^\mathrm{L}_1$ given by Eqs.~(\ref{eq:a-27a})--(\ref{eq:a-27c})
for $\sigma \ll 1$. The details of the derivation are given in
Appendix~\ref{app:AnalyticRBF}. The results are
\begin{align}
  \label{eq:3-12a}
  b^\mathrm{L}_1
  &\simeq \frac{2^8\sqrt{2}}{3^2\cdot 7\sqrt{\pi}\,\sigma} \simeq
    \frac{3.242}{\sigma} ,
  \\
  \label{eq:3-12b}
  b^\mathrm{L}_2
  &\simeq \frac{10}{\sigma^2},
  \\
  \label{eq:3-12c}
  \omega^\mathrm{L}_1
  &\simeq -\frac{5}{2 \sigma^2}.
\end{align}
Comparing these expressions with Fig.~\ref{fig:coeffwo} in
Appendix~\ref{app:AnalyticRBF}, the power-law behaviors of the bias
coefficients for $\sigma \lesssim 0.1$ are accurately explained by the
above asymptotic formula.

\subsection{\label{subsec:initPS} Initial PBH power spectrum with
  primordial non-Gaussianity }

Equations~(\ref{eq:3-8a}) and (\ref{eq:3-8b}) are the renormalized
bias functions that we need for evaluating the effects of primordial
non-Gaussianity in the initial PBH power spectrum at the lowest order.
From Eq.~(\ref{eq:3-8b}), we have
$c_2(\bm{p},-\bm{p}) = (b^\mathrm{L}_2 +
2\omega^\mathrm{L}_1)W^2(pR)$. Substituting this form into
Eq.~(\ref{eq:2-13}), the integral $C_2$ is calculated to be
\begin{equation}
  \label{eq:3-13}
  C_2 = \left(b^\mathrm{L}_2 + 2\omega^\mathrm{L}_1\right)\sigma^2.
\end{equation}
Thereby, Eq.~(\ref{eq:2-12}) reduces to
\begin{equation}
  \label{eq:3-14}
  P_\mathrm{PBH}(k)
  \simeq
    \tau_\mathrm{NL} \left(b^\mathrm{L}_2 + 2\omega^\mathrm{L}_1\right)^2 \sigma^4
    \frac{P_\mathrm{L}(k)}{\mathcal{M}^2(k)}.
\end{equation}
This is a main result of this paper.
\begin{figure}
  \centering
  \includegraphics[height=12.5pc]{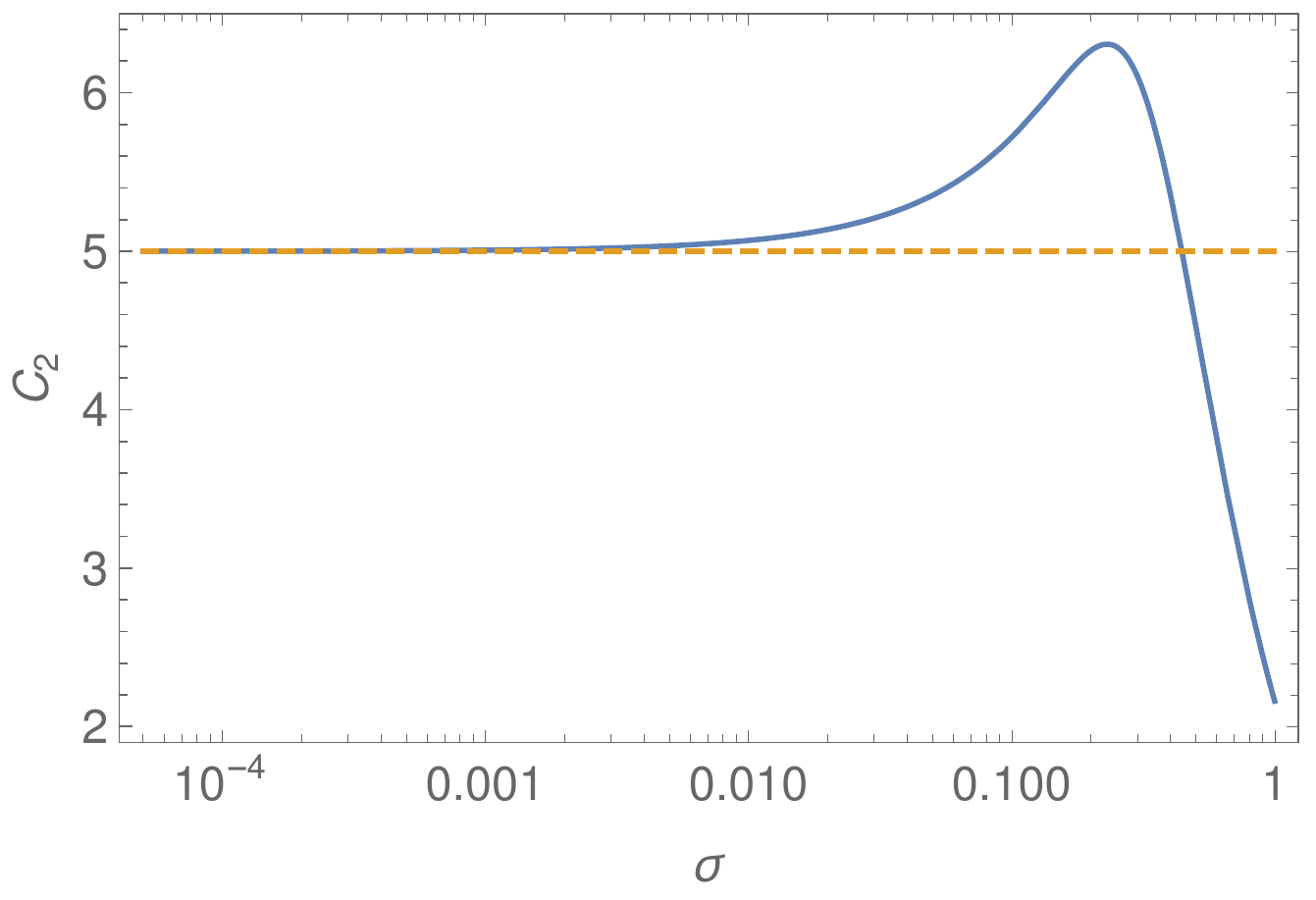}
  \caption{\label{fig:C2wo} The integral $C_2$ of Eq.~(\ref{eq:3-13}) in
  the model of Ref.~\cite{Har16}. Blue solid line: the result of the
  numerical integrations. Orange dashed line: asymptotic formula $C_2=5$
  for $\sigma\ll 1$.}
\end{figure}
In Fig.~\ref{fig:C2wo}, the result of the numerical integrations for
$C_2$ (without effects of angular momentum) is plotted. In the case of
$\sigma \ll 1$, substituting Eqs.~(\ref{eq:3-12b}) and
(\ref{eq:3-12c}) into Eq.~(\ref{eq:3-13}) gives $C_2 \simeq 5$, and we
have
\begin{equation}
  \label{eq:3-15}
  P_\mathrm{PBH}(k)
  \simeq
    25 \tau_\mathrm{NL} \frac{P_\mathrm{L}(k)}{\mathcal{M}^2(k)} = 25
    \tau_\mathrm{NL} P_\mathcal{R}(k). 
\end{equation}
Interestingly, the analytic estimate of Eq.~(\ref{eq:3-15})
corresponds to the formula of high-peak limit, Eq.~(\ref{eq:2-31})
with $\nu = \sqrt{5}$. However this does not imply the PBH formation
at the MD epoch corresponds to the density peaks of this height
because $b^\mathrm{L}_1 \ne \sqrt{5}/\sigma$. The asymptotic formula
of Eq.~(\ref{eq:3-15}) is accurately applicable for
$\sigma \lesssim 0.01$, as one can see from Fig.~\ref{fig:C2wo}. If
only the 10\% accuracy is required, the same formula is applicable for
$\sigma \lesssim 0.1$.

\subsection{\label{subsec:AngularMomentum}
  Effects of angular momentum
}

In Ref.~\cite{Har17}, the model of Ref.~\cite{Har16} is extended to
include the effect of rotation, which turns out to play important
roles in the formation of PBHs. The effect of angular momentum in the
formation of PBH in the MD epoch exponentially suppresses the
amplitude of $\beta_0$ for small values of $\sigma$ \cite{Har17}.
According to Ref.~\cite{Har17}, the effect of the angular momentum can
be taken into account by changing the number density of PBH of
Eq.~(\ref{eq:3-4}) to
\begin{multline}
  \label{eq:3-41}
  n_\mathrm{PBH}(\alpha,\beta,\gamma)\\ =
  \frac{\bar{n}_\mathrm{PBH}}{\beta_0'} \Theta(\alpha)
  \Theta\left(\alpha + \beta + \gamma - \delta_\mathrm{th}\right)
  \Theta\left[1-h(\alpha,\beta,\gamma)\right],
\end{multline}
where
\begin{multline}
  \label{eq:3-42}
  \beta_0' = \int_0^\infty d\alpha \int_{-\infty}^\alpha d\beta
   \int_{-\infty}^\beta d\gamma
  \Theta\left(\alpha + \beta + \gamma - \delta_\mathrm{th}\right)
  \\ \times
  \Theta\left[1-h(\alpha,\beta,\gamma)\right]
   w(\alpha,\beta,\gamma),
\end{multline}
and $\delta_\mathrm{th}$ is the density threshold above which the Kerr
bound $L\leq GM^2/c$ is satisfied, where $L$ and $M$ are the angular
momentum and mass of the black hole, respectively. This bound is
required in order to have a black hole at the center of the Kerr
metric.

There is an ambiguity in the model on initial quadrupole moment of the
mass, which is parametrized by $q$ in Ref.~\cite{Har17}. There are two
cases which are considered in this reference,
\begin{equation}
  \label{eq:3-43}
  \delta_\mathrm{th(1)} \equiv \frac{12}{125} q^2, \quad
  \delta_\mathrm{th(2)} \equiv \left(\frac{2}{5}
    \mathcal{I} \sigma \right)^{2/3},
\end{equation}
where $\mathcal{I}$ is another parameter of order unity which
characterizes the variance of angular momentum (see Ref.~\cite{Har17}
for explicit definitions of parameters $q$ and $\mathcal{I}$). In the
following calculation, we assume $q=\sqrt{2}$ and $\mathcal{I}=1$ to
match Fig.~5 of Ref.~\cite{Har17}. We ignore the effect of the finite
duration of the MD epoch. The two thresholds, $\delta_\mathrm{th(1)}$
and $\delta_\mathrm{th(2)}$, are called first and second order,
respectively. In Ref.~\cite{Har17}, it is suggested that the
second-order case is relatively realistic in practice. In
Eqs.~(\ref{eq:3-41}) and (\ref{eq:3-42}), the extra factor
$\Theta(\alpha + \beta + \gamma - \delta_\mathrm{th})$ is inserted in
the integrals of Eqs.~(\ref{eq:3-4}) and (\ref{eq:3-5}). The numerical
calculations of the bias parameters $b^\mathrm{L}_1$, $b^\mathrm{L}_2$
and $\omega^\mathrm{L}_1$ are similarly possible as in the case of
previous subsections. In practice, the function $z_*(t,u)$ in
Eqs.~(\ref{eq:b-5}) and (\ref{eq:b-7}) is substituted by
$z_*(t,u) \rightarrow z_0(t,u) \equiv \mathrm{max}[z_*(t,u),
z_\mathrm{th}(t)]$, where
$z_\mathrm{th} \equiv \delta_\mathrm{th}/(3t)$.

\begin{figure}
\centering
  \includegraphics[height=12.5pc]{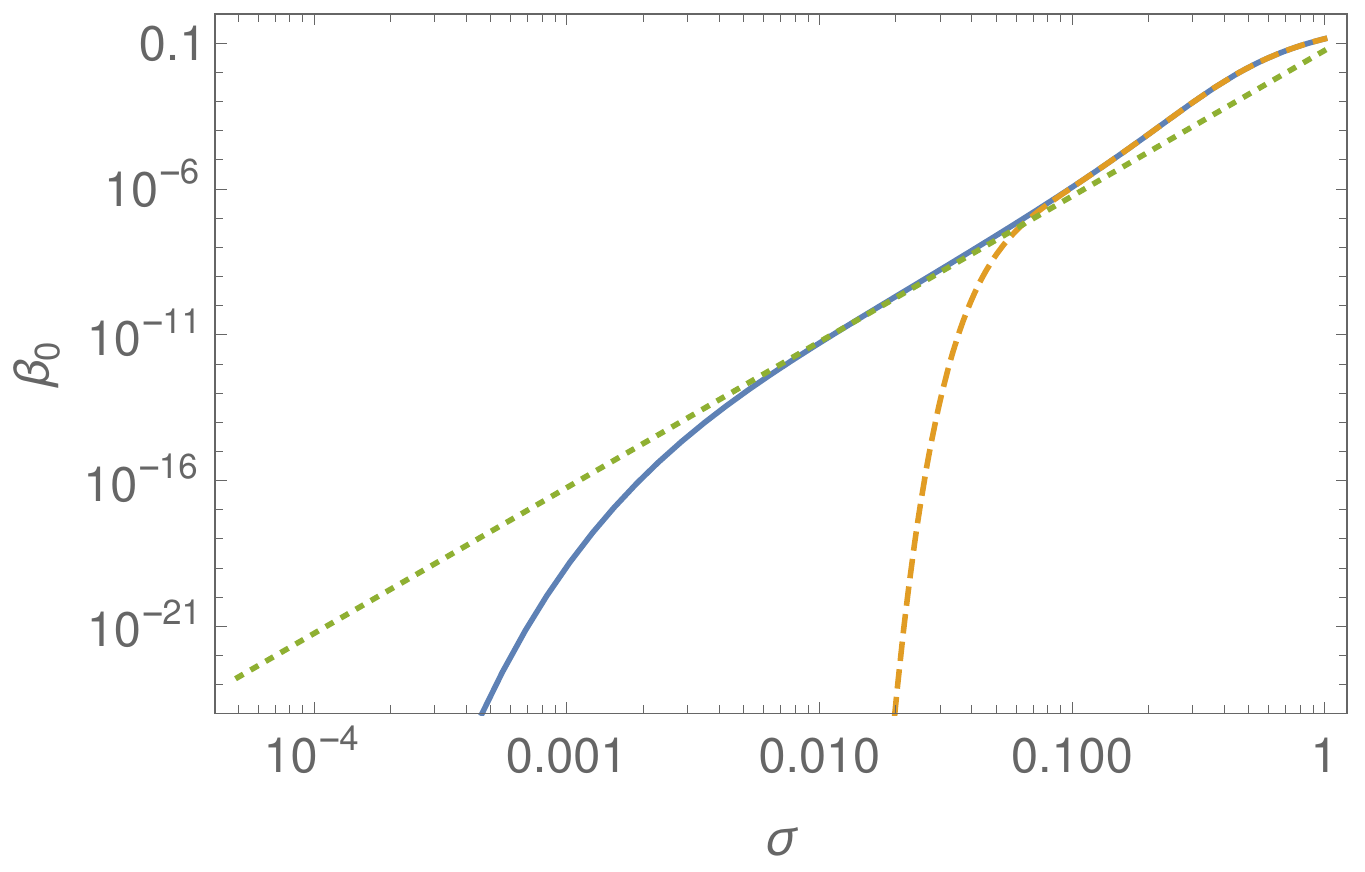}
\caption{\label{fig:beta0wa} The production probability of PBH with
  effects of angular momentum in the model of Ref.~\cite{Har17}. Blue
  solid line: second-order case with $\delta_\mathrm{th(2)}$. Orange
  dashed line: first-order case with $\delta_\mathrm{th(1)}$. Green
  dotted line: the asymptotic formula of $\sigma \ll 1$ without the
  effects of angular momentum. }
\end{figure}
In Fig.~\ref{fig:beta0wa}, the production probability of PBH with
effects of angular momentum is plotted. This figure reproduces the
corresponding result of Fig.~5 of Ref.~\cite{Har17}. The second-order
case is approximately described by the asymptotic formula without the
effects of angular momentum in $0.005 \lesssim \sigma \lesssim 0.1$.

\begin{figure}
\centering
  \includegraphics[height=12.5pc]{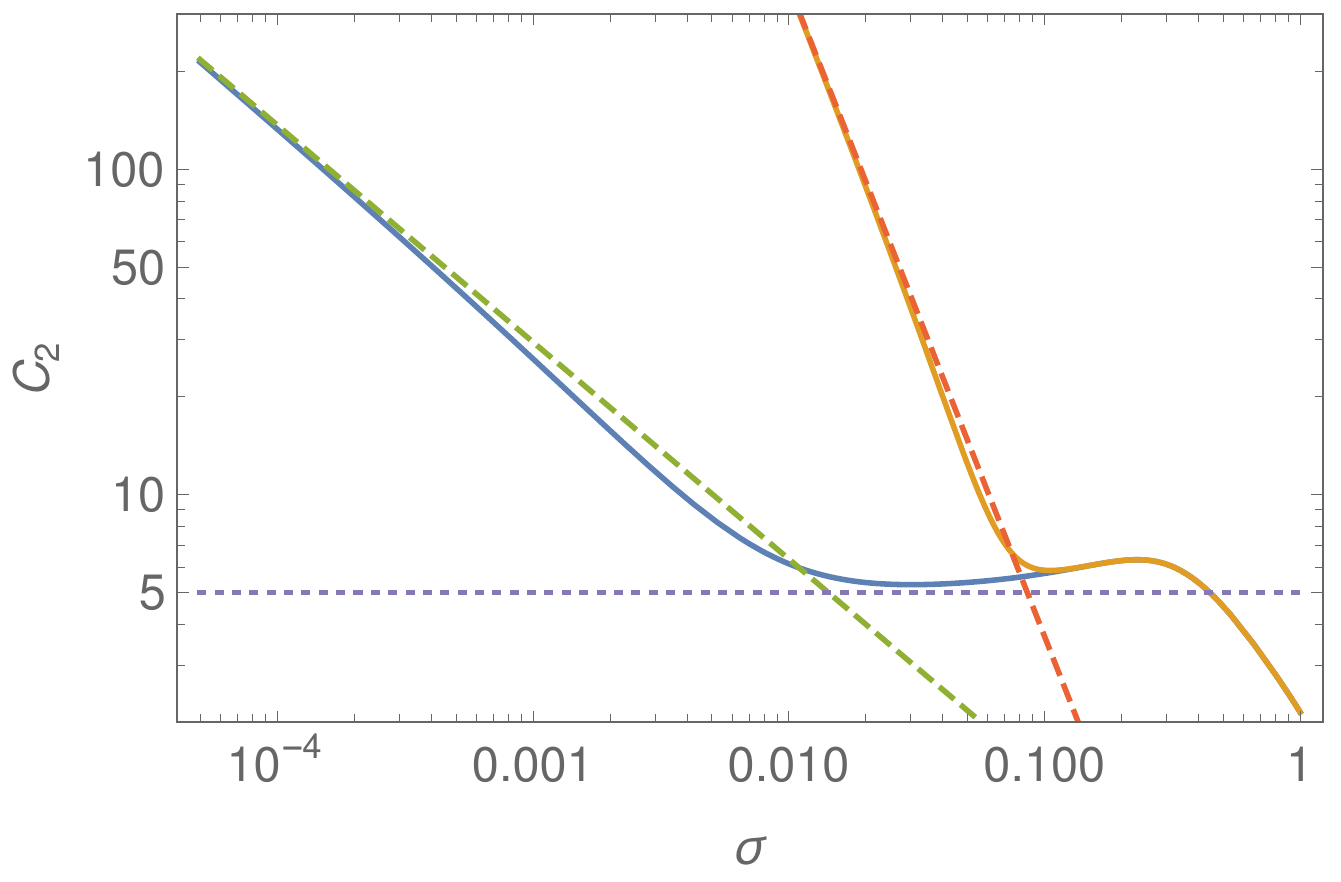}
\caption{\label{fig:C2wa} The integral $C_2$ of Eq.~(\ref{eq:3-13}) in
  the model of Ref.~\cite{Har17}. Blue solid line: the result of the
  numerical integrations in the second-order case. Green dashed line:
  asymptotic formula for $\sigma \ll 1$ in the second-order case.
  Orange solid line: the result in the first-order case. Orange dashed
  line: asymptotic formula for $\sigma \ll 1$ in the first-order case.
  Purple dotted line: asymptotic formula $C_2=5$ for $\sigma\ll 1$
  without the effects of angular momentum.}
\end{figure}
In Fig.~\ref{fig:C2wa}, the result of the numerical integrations for
$C_2$ with the effects of angular momentum is plotted. Comparing it
with Fig.~\ref{fig:C2wo}, the effects of angular momentum are
significant in $\sigma \lesssim 0.01$ for the second-order case and
$\sigma \lesssim 0.1$ for the first-order case.
Substituting the calculated values of $C_2$ into Eq.~(\ref{eq:2-12}),
we obtain the estimate of $P_\mathrm{PBH}(k)$ with the effects of
angular momentum.

The behaviors of the parameters $b^\mathrm{L}_1$, $b^\mathrm{L}_2$,
and $\omega^\mathrm{L}_1$ in $\sigma \ll\delta_\mathrm{th} \ll 1$ can
also be explained by considering the asymptotic limit of the
integrals. They are given in the second subsection of
Appendix~\ref{app:AnalyticRBF}, and the results are
\begin{align}
  \label{eq:3-44a}
  b^\mathrm{L}_1
  &\simeq \frac{\delta_\mathrm{th}}{\sigma^2},
  \\
  \label{eq:3-44b}
  b^\mathrm{L}_2
  &\simeq \frac{{\delta_\mathrm{th}}^2}{\sigma^4},
  \\
  \label{eq:3-44c}
  \omega^\mathrm{L}_1
  &\simeq 0.03400
  \frac{{\delta_\mathrm{th}}^4}{\sigma^4}  - \frac{5}{2\sigma^2}.
\end{align}
These asymptotic formulas explain the results of numerical integration
for $\sigma \ll 1$ fairly well. From the above, Eq.~(\ref{eq:3-13}) in
the limit of $\sigma \ll\delta_\mathrm{th} \ll 1$ is dominated by
$b^\mathrm{L}_2$ and is given by
\begin{equation}
  \label{eq:3-45}
  C_2 \simeq \frac{{\delta_\mathrm{th}}^2}{\sigma^2}.
\end{equation}
In the regime where the above approximation applies, we have
\begin{equation}
  \label{eq:3-46}
  P_\mathrm{PBH}(k) \simeq \tau_\mathrm{NL}
  \left(\frac{\delta_\mathrm{th}}{\sigma}\right)^4
  \frac{P_\mathrm{L}(k)}{\mathcal{M}^2(k)} =\tau_\mathrm{NL}
  \left(\frac{\delta_\mathrm{th}}{\sigma}\right)^4
  P_\mathcal{R} (k).  
\end{equation}
Identifying $\delta_{\rm th} / \sigma = \nu$, the above expression is
similar to the formula of the high-peak limit, Eq.~(\ref{eq:2-31}),
although $\delta_\mathrm{th}$ generally depends on $\sigma$ in this
case.

\begin{figure}
\centering
  \includegraphics[height=12.5pc]{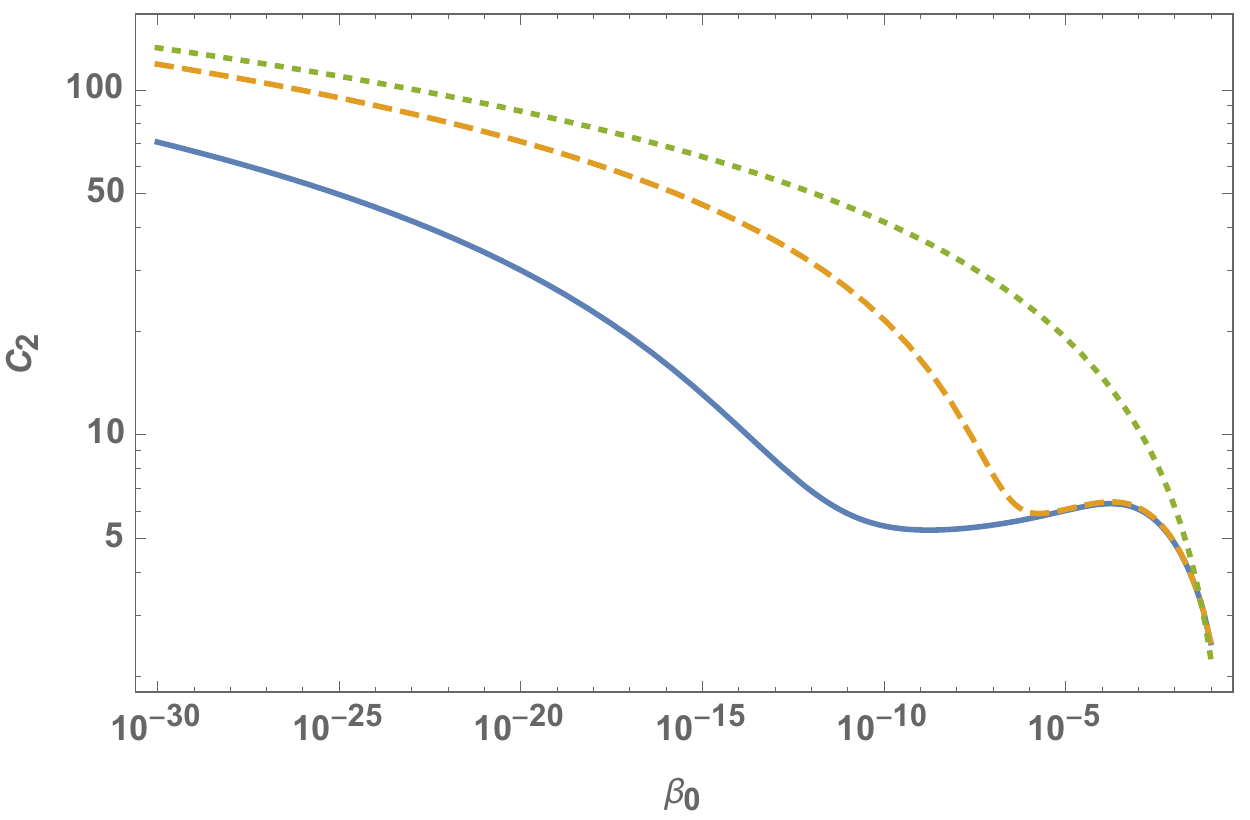}
  \caption{\label{fig:C2vsBeta} The integral $C_2$ of
    Eq.~(\ref{eq:3-13}) in the model of Ref.~\cite{Har17} as a
    function of $\beta_0$. The blue solid line is for the second-order
    case. The orange dashed line corresponds to the first-order case.
    The green dotted line represents the case for the PBH formation in
    the RD epoch calculated by Eqs.~(\ref{eq:2-28}) and
    (\ref{eq:2-29-1}).}
\end{figure}

One should note that the production probability $\beta_0$ is
exponentially suppressed in this regime, and the number density of
PBHs is extremely small when the above approximation applies. In fact,
$\beta_0$ is required to be roughly
$\mathcal{O}(10^{-15})$--$\mathcal{O}(10^{-10})$ depending on the mass
of PBHs \cite{Carr:2009jm}\footnote{ \label{fn:f-beta_relation}
  Reference~\cite{Carr:2009jm} focused on the PBHs formed during a RD
  epoch, in which $\Omega_\text{PBH} \sim a^1$, and hence one should
  be careful when applying the result in Ref.~\cite{Carr:2009jm} to
  PBHs formed in a MD epoch, in which $\Omega_\text{PBH} \sim a^0$.}
in order for PBHs to be a relevant component in dark matter (DM). As
can be seen in Fig.~\ref{fig:C2vsBeta} in the second-order case (blue
solid line), $C_2$ is about $5$--$10$ for the above range of
production probability $\beta_0$. As we have mentioned above, there is
an ambiguity for taking the effect of the angular momentum into
account. Since the PBH power spectrum is proportional to ${C_2}^2$, we
consider $C_2 = 5$ for the PBHs formed in the MD epoch as a
conservative value for the amplitude of the power spectrum in the
following discussion. More precise discussion is given in
Appendix~\ref{app:constraints}, taking into account the dependence of
$C_2$ on $\sigma$ (or equivalently on $\beta_0$).

\section{\label{sec:Observational}
  Observational implications
}  

In the previous sections, we obtained the theoretical power spectrum
of PBHs which is formed in a MD epoch in the presence of local-type
non-Gaussianity. Whether or not this signal has any observable effect
is another issue, which we consider in this section. We first estimate
the effect of shot noise for possible candidates of PBHs which are
connected to observations. Next, we consider the isocurvature
fluctuations produced by the PBHs, which can place constraints on the
model by comparing with observations of the CMB.

\subsection{\label{subsec:ShotNoise}
  Shot-noise contribution
}

When the produced number of PBHs is too small, their actual power
spectrum does not necessarily follow the theoretical prediction
because of randomness in the position of each object, or the Poisson
shot noise effect. Before we conclude the PBH power spectrum estimated
in the previous section is physically meaningful, we have to compare
them with the power spectrum of shot noise.
The shot-noise contribution to the power spectrum is given by
\begin{align}
  P_\mathrm{SN}(k) = \frac{1}{\bar{n}_\mathrm{PBH}},
\end{align}
where $\bar{n}_\text{PBH}$ is the mean number density of PBHs
estimated to be
\begin{align}
  \bar{n}_\text{PBH}
  &= \frac{3 H_0^2}{8\pi G} \Omega_\text{CDM}
    \frac{f_\text{PBH}}{M_\text{PBH}}
    \nonumber \\
  &= 3 \times 10^{22}\, \textrm{Mpc}^{-3}\left(
    \frac{\Omega_\text{CDM}h^2}{0.12} \right)
    \left( \frac{f_\text{PBH}}{1} \right)
    \left( \frac{M_\text{PBH}}{10^{-12} M_\odot} \right)^{-1}.
\end{align}
Here $H_0 = 100\, h\,\textrm{km/s/Mpc}$ is the Hubble constant,
$\Omega_\text{CDM}$ is the energy density fraction of cold dark
matter, $f_\mathrm{PBH}$ is the relative ratio of the energy density
of PBHs to those of total dark matter, and $M_\text{PBH}$ is the mass
of PBHs.
Thus, the shot-noise contribution can be expressed in terms of
$f_{\rm PBH}$ and $M_{\rm PBH}$, as
\begin{align}
  P_\text{SN}
  &= 3\times 10^{-23}\, \textrm{Mpc}^3  \left(
    \frac{\Omega_\text{CDM}h^2}{0.12} \right)^{-1} \left(
    \frac{f_\text{PBH}}{1} \right)^{-1} \left(
    \frac{M_\text{PBH}}{10^{-12} M_\odot}
    \right). \label{eq:P_SN} 
\end{align}
Note that the magnitude of the shot noise is determined by the
combination $M_\text{PBH}/f_\text{PBH}$. This contribution behaves as
matter isocurvature fluctuations with blue tilt in terms of
dimensionless power spectrum
$\mathcal{P} (k) \equiv k^3 P (k)/(2\pi^2)$, and it would affect the
formation of structures on small scales. Based on this fact, one can
place a constraint on the abundance $f_{\rm PBH}$ of PBHs by using
observations of structures on small scales, such as Lyman-$\alpha$
forest~\cite{Afshordi:2003zb} and future 21cm
observations~\cite{Gong:2017sie}.

\begin{figure}
\centering
  \includegraphics[width=0.98\columnwidth]{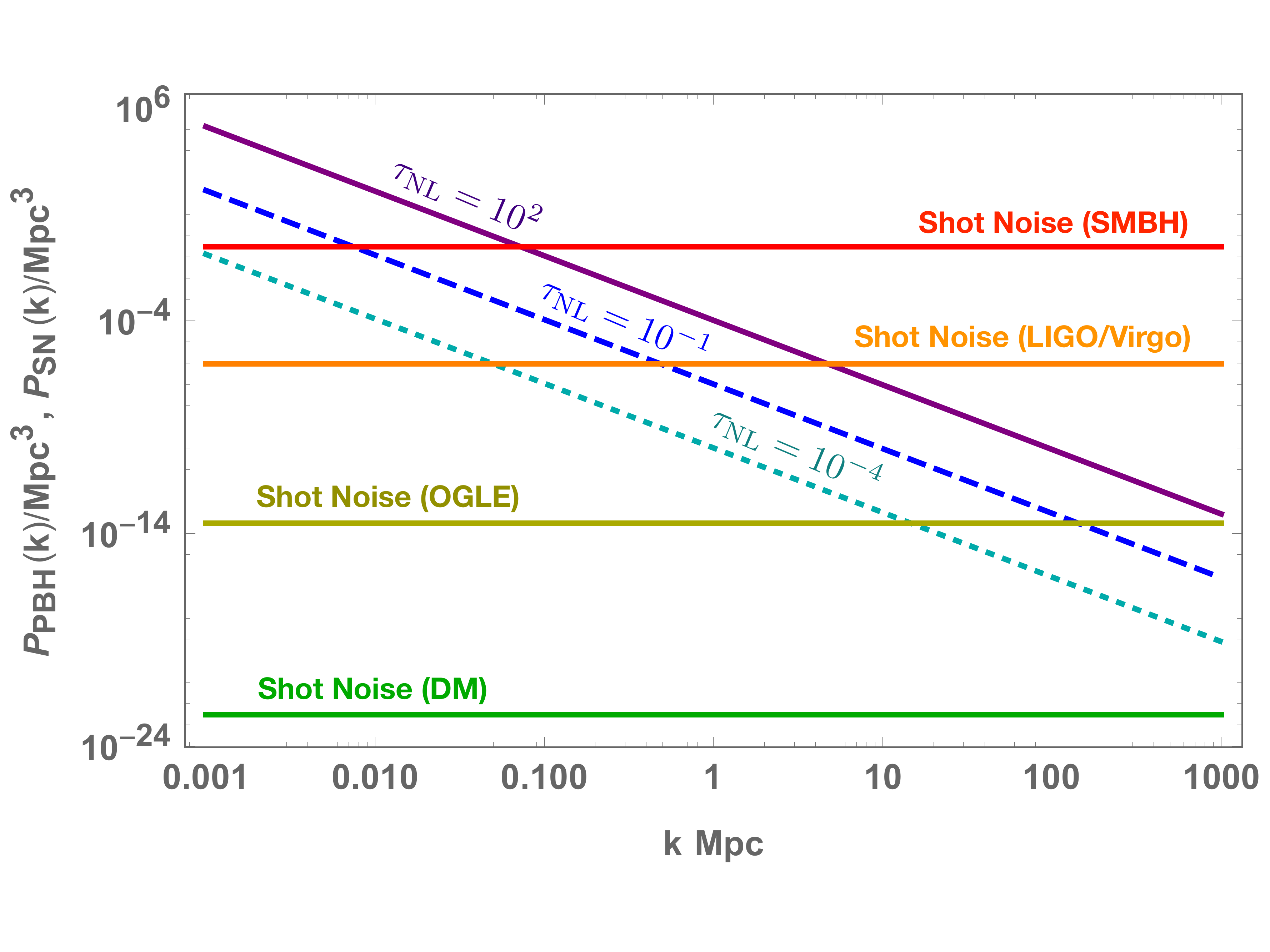}
  \caption{\label{fig:vsShotNoise} Comparison of the initial PBH power
    spectrum induced by primordial non-Gaussianity during the MD epoch
    [Eq.~\eqref{eq:3-15}] (oblique lines) with the shot noise
    contribution [Eq.~\eqref{eq:P_SN}] (horizontal lines) as a
    function of wave number $k$. The assumed typical values of
    $M_\text{PBH}/(M_\odot f_\text{PBH})$ for DM, OGLE, LIGO/Virgo,
    and SMBH are $10^{-12}$, $10^{-3}$, $10^{4.5}$, and $10^{10}$,
    respectively. Note that the SMBH case is irrelevant for PBH
    production in a MD epoch, but it is nevertheless shown because the
    figure is applicable to the case of the RD epoch (high-peak limit)
    by the replacement of the normalization of the oblique lines,
    $25 \to \nu^4$. }
\end{figure}

In Fig.~\ref{fig:vsShotNoise}, the shot-noise contributions given by
Eq.~\eqref{eq:P_SN} are compared with initial PBH power spectra with
the primordial non-Gaussianity given by Eq.~\eqref{eq:3-15}. Even
though the assumed value $C_2=5$ corresponds to the asymptotic value
without the effects of angular momentum, this gives the lower limit of
the power spectrum with the effects of angular momentum, since
$C_2 >5$ in the latter case. Here we assume
$\mathcal{P}_\mathcal{R}(k) = A_\mathrm{s} k^{n_\mathrm{s}-1}$ with
$A_\mathrm{s} =2.101 \times 10^{-9}$ and $n_\mathrm{s} = 0.9649$
\cite{Akrami:2018odb} for $k \lesssim 10^3$ Mpc$^{-1}$. The initial
PBH power spectrum only depends on the value of $\tau_{\rm NL}$ as
seen from Eq.~\eqref{eq:3-15}. In this figure, we show the power
spectrum with multiple choices of
$\tau_\text{NL} = 10^2 \text{ (purple) }, 10^{-1} \text{ (blue) },
\text{ and } 10^{-4} \text{ (dark cyan) }$. For the shot-noise
contributions, we consider typical values of
$M_\text{PBH}/f_\text{PBH}=10^{-12}M_\odot$, $10^{-3}M_\odot$,
$10^{4.5}M_\odot$, and $10^{10}M_\odot$. These values correspond to
the cases of all the dark matter (DM: light green,
$M_{\rm PBH} = 10^{-12} M_\odot$,
$f_{\rm PBH} = 1$)~\cite{Bartolo:2018rku}, excess events of OGLE
observations (OGLE: dark yellow, $M_{\rm PBH} = 10^{-5} M_\odot$,
$f_{\rm PBH} = 10^{-2}$)~\cite{Niikura:2019kqi, Tada:2019amh,
  Fu:2019ttf}, the origin of binary black holes leading to the
gravitational-wave events of LIGO/Virgo (LIGO/Virgo: orange,
$M_{\rm PBH} = 10^{1.5} M_\odot$,
$f_{\rm PBH} =
10^{-3}$)~\cite{Bird:2016dcv,Sasaki:2016jop,Sasaki:2018dmp}, and the
seeds of supermassive black holes (SMBH: red,
$M_{\rm PBH} = 10^{3.5} M_\odot$,
$f_{\rm PBH} =
10^{-6.5}$)~\cite{Kawasaki:2012kn,Kohri2014SMBH,Kawasaki:2019iis},
respectively. These are just benchmark points, and the actual allowed
region of $M_\mathrm{PBH}$ is not a point but a band. Also, the
favored region for $M_\text{PBH}$ and $f_\text{PBH}$ has a large
uncertainty. However, we are not particularly interested in these
issues here.

Note also that for the PBH production in the early MD epoch, the SMBH
case requires a reheating temperature too low to be consistent with
big bang nucleosynthesis \cite{Hasegawa:2019jsa} (see also
Refs.~\cite{Kawasaki:1999na,Kawasaki:2000en,Hannestad:2004px,Ichikawa:2005vw,deSalas:2015glj}).
The reason why we nevertheless show the typical line corresponding to
this case in Fig.~\ref{fig:vsShotNoise} is because these results are
also applicable to the case of PBH production in the RD epoch after
making the replacement $25 \to \nu^4$ for the magnitude of the PBH
power spectrum (the high-peak limit is assumed). When the mass of PBHs
is less than $10^{7}\,M_\odot$, the threshold value for the production
of PBHs in the RD epoch is given by $\nu^4 \gtrsim 400$~\cite{TY15}.
In this case, the amplitude of the PBH power spectrum in
Fig.~\ref{fig:vsShotNoise} is 16 times larger than the plotted lines.

In Fig.~\ref{fig:vsShotNoise}, we see that the shot-noise contribution
becomes relatively unimportant on large scales because of the scale
dependence of the initial PBH power spectrum approximately
$k^{n_\mathrm{s}-3}$. Also, for DM or OGLE, the shot noise is
completely negligible for the scales of CMB and the large-scale
structure.

\subsection{\label{subsec:Isocurvature}
  Constraints from isocurvature mode in CMB
}

The initial clustering of primordial black holes induced from the
primordial non-Gaussianity would be observed as isocurvature
perturbations. The isocurvature perturbations are well constrained by
CMB, and thus the abundance of PBHs $f_\mathrm{PBH}$ or the magnitude
of $\tau_\mathrm{NL}$ is constrained as well.

The PBH isocurvature perturbations are given by
\begin{equation}
  \mathcal{I}_{\rm PBH} =
    \delta_\mathrm{PBH} - \frac{\delta}{1+w},
    \label{eq:Iso}
\end{equation}
where $\delta_\mathrm{PBH}$ is the density contrast of PBHs, $\delta$
is the density contrast of the dominant component of the Universe
which turns into the radiation component in the RD epoch, and $w$ is
the equation-of-state parameter of the latter component. On comoving
slices, from Eq.~\eqref{eq:2-3}, the density contrast of the dominant
component of the Universe must be much suppressed by $k^2$ in
$\mathcal{M}(k)$ in the large-scale limit. Here, we consider the PBH
isocurvature perturbations at CMB scales which are much larger than
the PBH formation scale, and hence the PBH isocurvature perturbations
are simply given by $\mathcal{I}_{\rm PBH} = \delta_\mathrm{PBH}$
where $\delta$ is negligible on large scales.
    
The Planck Collaboration gives a constraint on the total CDM
isocurvature perturbations, and thus, if the PBHs exist as a DM
component with the fraction, $f_{\rm PBH}$, the power spectrum of the
total CDM isocurvature perturbations can be given as
\begin{equation}
    \label{eq:iso1}
    P_{\mathcal{I}\mathcal{I}}(k) = f_\mathrm{PBH}^2
        P_\mathrm{PBH}(k) ,
\end{equation}
where we have assumed that the other DM components do not have any
isocurvature perturbations.
Substituting Eq.~\eqref{eq:2-12} into the above expression, the power
spectrum of CDM isocurvature perturbation is given by
\begin{equation}
    P_{\mathcal{I}\mathcal{I}} (k) = 
    {C_2}^2 {f_\mathrm{PBH}}^2
    \tau_\mathrm{NL} P_\mathcal{R}(k) .
  \label{eq:iso2}
\end{equation}
The above equation holds for PBH formation in both the RD
($C_2=\nu^2$) \cite{TY15} and MD ($C_2 \geq 5$) epochs.

The constraint on the CDM isocurvature perturbations (for correlated
case) placed by Planck 2018 \cite{Akrami:2018odb} is given by
$P_{\mathcal{II}}/P_{\mathcal{R}} \lesssim 10^{-3}$ on CMB scales.
Applying this constraint to our result, we obtain an upper bound on
$f_\text{PBH}$ depending on the value of $\tau_{\rm NL}$ as follows:
\begin{align}
  f_\text{PBH} < \frac{3 \times 10^{-2}}{C_2
  \sqrt{\tau_\text{NL}}}.
  \label{eq:f_upper-bound} 
\end{align}
In the limit of $\tau_\text{NL} \to 0$, the isocurvature upper bound
on $f_\text{PBH}$ disappears, which is consistent with the
understanding that it is non-Gaussianity that induces the PBH
isocurvature perturbations.

\begin{figure}[t] 
\centering
  \includegraphics[width=0.98\columnwidth]{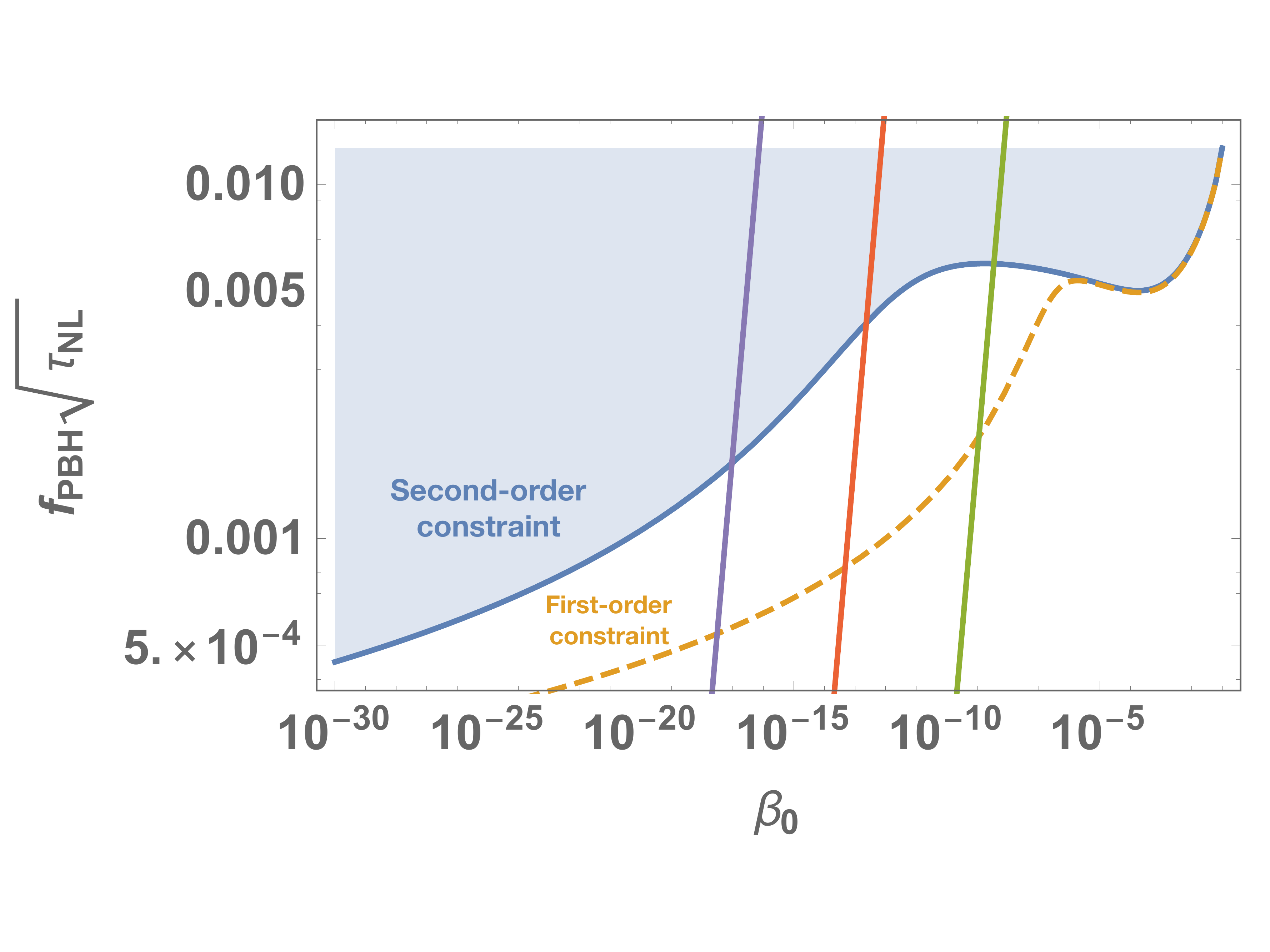}
  \caption{\label{fig:max_fSqrtTau_MD} Upper bounds on
    $f_\text{PBH} \sqrt{\tau_\text{NL}}$ as a function of $\beta_0$ in
    the case of PBH production in a MD epoch by the second-order
    mechanism (blue solid line) and by the first-order mechanism
    (orange dashed line). Three examples of
    $f_\text{PBH} \sqrt{\tau_\text{NL}}$ are plotted as functions of
    $\beta_0$: $T_\text{R}= 10^{-2}$ GeV (Green line), $10^{2}$ GeV (red
    line), and $10^6$ GeV (purple line) with $\tau_\text{NL} = 10^{-2}$
    for all cases. }
\end{figure}

The above constraint can be rewritten as
$f_\text{PBH} \sqrt{\tau_\text{NL}} < 3\times 10^{-2}/C_2$, which is
an upper bound on the combination $f_\text{PBH} \sqrt{\tau_\text{NL}}$
given the value of $C_2$, which in turn depends on $\beta_0$. This is
shown in Fig.~\ref{fig:max_fSqrtTau_MD} in the case of PBH production
in a MD epoch. The shaded region is excluded by the isocurvature
constraint (the blue line corresponding to the second-order case). The
first-order constraint is also shown by the orange dashed line.
Actually, $f_\text{PBH}$ (present abundance) itself depends linearly
on $\beta_0$ (initial abundance) (see, e.g.,
Refs.~\cite{Inomata:2016rbd, Kohri:2018qtx}),
\begin{align}
  f_\text{PBH}
  &\equiv \frac{\rho_\text{PBH}}{\rho_\text{CDM}} \nonumber \\
  &= \frac{g_*(T)g_{*,s}(T_\text{eq})T}{g_*(T_\text{eq})g_{*,s}(T)
    T_\text{eq}} \gamma \beta_0
    \frac{\Omega_\text{m}}{\Omega_\text{CDM}},
    \label{eq:f-beta}
\end{align}
where $g_* (T)$ and $g_{*,s} (T)$ are the effective relativistic
degrees of freedom for energy density and entropy\footnote{We use
  precise functions of $g_* (T)$ and $g_{*,s} (T)$ provided by
  Ref.~\cite{Saikawa:2018rcs}.}, respectively; $T_\text{eq}$ is the
temperature at the matter-radiation equality; $\gamma$ ($\approx 0.2$
in the RD epoch~\cite{Carr:1975qj}) is an efficiency parameter
parametrizing how much fraction of the horizon mass goes into the PBH;
and $\Omega_\text{m}$ is the energy density fraction of total matter.
The above expression should be evaluated at the temperature when the
scale that becomes PBHs enters the horizon in the case of PBH
production in the RD epoch or at the reheating temperature in the case
of PBH production in the early MD epoch \cite{Carr:2017edp,
  Kohri:2018qtx}. The efficiency parameter in the early MD epoch has
some uncertainty, and we assume $\gamma=1$ following
Ref.~\cite{Kohri:2018qtx}. Using this relation, example lines of
$f_\text{PBH} \sqrt{\tau_\text{NL}}$ are shown in the same figure for
$T_\text{R}=10^{-2}$~GeV (green line), $10^2$~GeV (red line), and
$10^6$~GeV (purple line) with $\tau_\text{NL}=10^{-2}$. The
intersection of such a line and the constraint curve gives the upper
bound on $f_\text{PBH}$ given the values of $T_\text{R}$ and
$\tau_\text{NL}$. These lines are drawn from numerical solutions as
explained in Appendix~\ref{app:constraints}.

Unless $\sigma$ is too large, the value of $C_2$ is always larger than
$5$ regardless of uncertainty in estimating the effect of angular
momentum. In the case of PBH formation in the RD epoch, the value of
$C_2$ is much larger than $5$. Therefore, one can conclude that the
bound $C_2 \geq 5$ is conservatively satisfied in the PBH formation in
both the RD and MD epochs. Combining this bound with
Eq.~(\ref{eq:f_upper-bound}), we have a conservative upper bound,
\begin{equation}
  \label{eq:consup}
  f_\mathrm{PBH} \sqrt{\tau_\mathrm{NL}} < 6 \times 10^{-3}.
\end{equation}
If we only consider the PBH formation in the RD epoch,
$C_2\gtrsim 20$, the upper bound becomes smaller,
$f_\mathrm{PBH} \sqrt{\tau_\mathrm{NL}} < 2 \times 10^{-3}$. In either
case, the order of magnitude of the upper bounds is not significantly different.

It is remarkable that the possibility of PBHs being all of the dark
matter ($f_\mathrm{PBH}=1$) can be excluded by the isocurvature
constraint for $\tau_\text{NL} \gtrsim 4 \times 10^{-5}$ (MD,
$C_2 \geq 5$) and $\tau_\text{NL} \gtrsim 3 \times 10^{-6}$ (RD,
$C_2 \gtrsim 20$). In other words, the hypothesis of all the dark
matter places a conservative upper bound on $\tau_\text{NL}$,
irrespective of the formation epoch,
\begin{align}
  \tau_\text{NL} < 4 \times 10^{-5}
  & &  (100\%~\text{PBH dark matter}). 
\end{align}

\section{\label{sec:Conclusions}
  Conclusions
}

In this paper, we derive a general prediction of the initial
clustering of PBHs in the presence of the parameter $\tau_\mathrm{NL}$
of local-type primordial non-Gaussianity. Using the formalism of iPT,
we generally have Eq.~(\ref{eq:2-12}), which is the prediction of the
large-scale power spectrum in the presence of $\tau_\mathrm{NL}$. In
the case of PBHs, the result is given by
$P_\mathrm{PBH}(k) \simeq \tau_\mathrm{NL} {C_2}^2 P_\mathcal{R}(k)$
for $k \ll R^{-1}$, where $R$ is the initial radius of proto-PBHs with
mass $M \sim 4\pi R^3/3$ up to the efficiency parameter. Evaluating
the integral $C_2$ in the high-peak limit, the linear power spectrum
of PBH formed in the RD epoch is given by Eq.~(\ref{eq:2-31}), which
is consistent to the previous work \cite{TY15,SY19}. In the case of
PBHs formed in a MD epoch, we adopt a model of
Refs.~\cite{Har16,Har17} and evaluate the initial power spectrum of
PBHs.

The integral $C_2$ is a decisive factor for the amplitude of initial
clustering of PBHs in the presence of $\tau_\mathrm{NL}$. For the
thresholded regions, it is given by $C_2 = H_1(\nu)/H_{-1}(\nu)$, and
this result reduces to $C_2 = \nu^2$ in the high-peak limit
$\nu \gg 1$. In the case of PBHs formed in a MD epoch, the integral
can be numerically evaluated. In the regime where the effects of
angular momentum are neglected, we have an analytic estimate
$C_2 \simeq 5$ for $\sigma \ll 1$. In the regime where the angular
momentum is important, we have another analytic estimate
$C_2 \simeq (\delta_\mathrm{th}/\sigma)^2$, where $\delta_\mathrm{th}$
is given by Eq.~(\ref{eq:3-43}). In general, one can evaluate the
value of $C_2$ by numerical integrations. The results are given in
Figs.~\ref{fig:C2wo} and \ref{fig:C2wa} without and with effects of
angular momentum, respectively.

Because of the approximately $k^{n_\mathrm{s}-4} \sim k^{-3}$ scaling
of the PBH power spectrum from the primordial non-Gaussianity, the
shot-noise contributions are relatively unimportant on large scales,
unless the mass of PBHs is extremely large and the number density of
PBHs is extremely small.

The clustering of PBHs produces the isocurvature perturbations in the
early Universe. The isocurvature power spectrum is proportional to
${C_2}^2 {f_\mathrm{PBH}}^2 \tau_\mathrm{NL}$. Putting $C_5 = 5$ as a
conservative value, the current constraint by the Planck satellite
gives an upper bound,
$f_\mathrm{PBH} \sqrt{\tau_\mathrm{NL}} < 6 \times 10^{-3}$. On one
hand, unless the non-Gaussianity parameter $\tau_\mathrm{NL}$ is
smaller than approximately $4\times 10^{-5}$, the hypothesis that all
the dark matter is made of PBHs is excluded. On the other hand, if all
the dark matter is made of PBHs, the parameter $\tau_\mathrm{NL}$ of
the primordial non-Gaussianity should be smaller than
$4 \times 10^{-4}$.

\begin{acknowledgments}
  This work was supported by JSPS KAKENHI Grants No.~JP16H03977
  (T.M.), No.~JP19K03835 (T.M.), No.~JP17H01131 (K.K.), and
  No.~JP17J00731 (T.T.); MEXT Grant-in-Aid for Scientific Research on
  Innovative Areas Grants No.~JP15H05889 (K.K.), No.~JP18H04594
  (K.K.), No.~JP19H05114 (K.K.), No.~JP15H05888 (S.Y.), and
  No.~JP18H04356 (S.Y.); Grant-in-Aid for JSPS Fellows (T.T.); and
  World Premier International Research Center Initiative, MEXT, Japan
  (K.K. and S.Y.).
\end{acknowledgments}

\newpage

\appendix
\onecolumngrid
\appendix
\onecolumngrid

\section{\label{app:RBF}
  Derivation of renormalized bias functions
}

In this Appendix, we derive the renormalized bias functions $c_1$ and
$c_2$ in our model of the number density $n_\mathrm{PBH}$. This number
density $n_\mathrm{PBH}$ is a function of a finite number of variables,
$\varphi_{ij}$. In this case, the renormalized bias function of
Eq.~(\ref{eq:3-7}) reduces to \cite{Mat11}
\begin{equation}
  \label{eq:a-8}
  c_n(\bm{k}_1,\ldots,\bm{k}_n) =
  \frac{1}{\bar{n}_\mathrm{PBH}}
  \sum_{a_1,\ldots,a_n}
  \left\langle
    \frac{\partial^nn_\mathrm{PBH}}
    {\partial y_{a_1}\cdots\partial y_{a_n}}
  \right\rangle
  U_{a_1}(\bm{k}_1)\cdots U_{a_n}(\bm{k}_n),
\end{equation}
where
\begin{align}
  \label{eq:a-9a}
  \left( y_a \right)
  & =
    \left( \varphi_{11},  \varphi_{22}, \varphi_{33}, \varphi_{12},
    \varphi_{23}, \varphi_{13} \right),
  \\                   
  \label{eq:a-9b}
  \left( U_a \right)
  & =
    W(kR) \left( {\hat{k}_1}^2, {\hat{k}_2}^2, {\hat{k}_3}^2,
    \hat{k}_1\hat{k}_2, \hat{k}_2\hat{k}_3, \hat{k}_1\hat{k}_3
    \right),
\end{align}
and $\hat{k}_i = k_i/k$ is the $i\mathrm{th}$ component of
$\bm{k}/k$. With the above definition, we have a relation,
$y_a(\bm{k}) = U_a(\bm{k})\delta_\mathrm{L}(\bm{k})$. We define an
operator,
\begin{equation}
  \label{eq:a-10}
  \mathcal{D}(\bm{k}) \equiv W(kR)
  \hat{k}_i \hat{k}_j \frac{\partial}{\partial\varphi_{ij}},
\end{equation}
where repeated indices are summed over, and partial derivatives are
taken as if $\varphi_{ij}$ and $\varphi_{ji}$ are independent
variables (because of the reason described in Ref.~\cite{MD16},
$n_\mathrm{PBH}$ can contain $\varphi_{ji}$ with $i>j$, provided that
$\varphi_{ji} = \varphi_{ij}$). With this operator,
Eq.~(\ref{eq:a-8}) reduces to \cite{MD16}
\begin{equation}
  \label{eq:a-11}
  c_n(\bm{k}_1,\ldots,\bm{k}_n) =
  \frac{1}{\bar{n}_\mathrm{PBH}}
  \left\langle
    \mathcal{D}(\bm{k}_1)\cdots\mathcal{D}(\bm{k}_n) n_\mathrm{PBH}
  \right\rangle =
  \frac{(-1)^n}{\bar{n}_\mathrm{PBH}}
  \int d^6\!y\,n_\mathrm{PBH}(\bm{y})
  \mathcal{D}(\bm{k}_1)\cdots\mathcal{D}(\bm{k}_n) \mathcal{P}(\bm{y}),
\end{equation}
where $\mathcal{P}(\bm{y})$ is a joint probability distribution
function of $\bm{y}=(y_a)$.

In the presence of initial non-Gaussianity, the probability
distribution function $\mathcal{P}$ is not strictly multivariate
Gaussian. However, as the lowest-order non-Gaussianity is concerned in
Eq.~(\ref{eq:2-12}), it is sufficient to use the renormalized bias
function derived from the Gaussian distribution function. Evaluation
of the renormalized bias functions of Eq.~(\ref{eq:a-11}) can be
performed in a method similar to that developed in Ref.~\cite{MD16}.
The Gaussian distribution function is given by
\begin{equation}
  \label{eq:a-12}
  \mathcal{P}(\bm{y}) =
  \frac{1}{\sqrt{(2\pi)^6 \det\mathcal{M}}}
  \exp\left(-\frac{1}{2}\bm{y}^\mathrm{T}\mathcal{M}^{-1}\bm{y}\right),
\end{equation}
where
\begin{equation}
  \label{eq:a-13}
  \mathcal{M}_{ab} =
  \left\langle y_a y_b \right\rangle =
  \int \frac{d^3\!k}{(2\pi)^3}
  U_a^*(\bm{k}) U_b(\bm{k}) P_\mathrm{L}(k),
\end{equation}
is the covariance matrix, and $P_\mathrm{L}(k)$ is the linear power
spectrum of the density perturbations. The elements of the covariance
matrix are given by
\begin{equation}
  \label{eq:a-13-1}
  \left\langle \varphi_{ij}\varphi_{kl} \right\rangle =
  \int \frac{d^3\!k}{(2\pi)^3}
  \hat{k}_i \hat{k}_j \hat{k}_k \hat{k}_l W^2(kR) P_\mathrm{L}(k)
  = \frac{\sigma^2}{15}
  \left(
    \delta_{ij}\delta_{kl} + \delta_{ik}\delta_{jl} +
    \delta_{il}\delta_{jk}
  \right) .
\end{equation}

The joint probability distribution function $\mathcal{P}(\bm{y})$
depends only on rotationally invariant quantities \cite{PGP09,GPP12}.
They are
\begin{equation}
  \label{eq:a-14}
  J_1 \equiv \frac{1}{\sigma}\varphi_{ii},\quad
  J_2 \equiv \frac{3}{2\sigma^2} \tilde{\varphi}_{ij} \tilde{\varphi}_{ji}, \quad
  J_3 = \frac{9}{2\sigma^3} \tilde{\varphi}_{ij} \tilde{\varphi}_{jk}
  \tilde{\varphi}_{ki},
\end{equation}
where
\begin{equation}
  \label{eq:a-15}
  \tilde{\varphi}_{ij} \equiv \varphi_{ij} - \frac{\sigma}{3} \delta_{ij} J_1 ,
\end{equation}
is the traceless part of $\varphi_{ij}$. With the invariant variables
of Eq.~(\ref{eq:a-14}), the distribution function of
Eq.~(\ref{eq:a-12}) reduces to \cite{GPP12}
\begin{equation}
  \label{eq:a-16}
  \mathcal{P}(\bm{y}) \propto
  \exp\left(-\frac{1}{2}{J_1}^2 - \frac{5}{2} J_2\right),
\end{equation}
up to the normalization factor.

Using relations,
\begin{equation}
  \label{eq:a-17}
  \frac{\partial J_1}{\partial \varphi_{ij}}
  = \frac{1}{\sigma}\delta_{ij}, \quad
  \frac{\partial J_2}{\partial \varphi_{ij}}
  = \frac{3}{\sigma^2}\tilde{\varphi}_{ji}, \quad
  \frac{\partial \tilde{\varphi}_{kl}}{\partial \varphi_{ij}}
  =\delta_{ik} \delta_{jl} - \frac{1}{3}\delta_{ij}\delta_{kl},
\end{equation}
the second-order derivatives are given by
\begin{align}
  \label{eq:a-18a}
  \frac{\partial}{\partial\varphi_{ij}} \mathcal{P}
  &= \left[
    \frac{\delta_{ij}}{\sigma} \frac{\partial}{\partial J_1} 
    + \frac{3 \tilde{\varphi}_{ij}}{\sigma^2} \frac{\partial}{\partial J_2}
    \right] \mathcal{P},
\\
  \label{eq:a-18b}
  \frac{\partial^2}{\partial\varphi_{ij}\partial\varphi_{kl}} \mathcal{P}
  &=
  \left[
  \frac{\delta_{ij} \delta_{kl}}{\sigma^2} \frac{\partial^2}{\partial {J_1}^2}
  - \frac{3}{\sigma^3} \left(
    \delta_{ij} \tilde{\varphi}_{kl} + \delta_{kl}
    \tilde{\varphi}_{ij} \right)
  \frac{\partial^2}{\partial J_1 \partial J_2} +
  \frac{9 \tilde{\varphi}_{ij} \tilde{\varphi}_{kl} }{\sigma^4}
  \frac{\partial^2}{\partial {J_2}^2}
  + \frac{3 \delta_{ik} \delta_{jl} - \delta_{ij} \delta_{kl}}{\sigma^2}
  \frac{\partial}{\partial J_2}
  \right] \mathcal{P}.
\end{align}
In calculating Eq.~(\ref{eq:a-11}), one notices that the number density
$n_\mathrm{PBH}$ and the distribution function $\mathcal{P}$ depend 
only on rotationally invariant variables. Thus we can first average
over the angular dependence in the product of operators
$\mathcal{D}(\bm{k})$. Denoting the angular average by
$\langle\cdots\rangle_\Omega$, Eq.~(\ref{eq:a-11}) reduces to
\begin{equation}
  \label{eq:a-19}
  c_n(\bm{k}_1,\ldots,\bm{k}_n) =
  \frac{(-1)^n}{\bar{n}_\mathrm{PBH}}
  \int d^6\!y\,n_\mathrm{PBH}(\bm{y})
  \left\langle
    \mathcal{D}(\bm{k}_1)\cdots\mathcal{D}(\bm{k}_n)
    \mathcal{P}(\bm{y})
  \right\rangle_\Omega.
\end{equation}
Using relations, 
\begin{equation}
  \label{eq:a-20}
  \left\langle \tilde{\varphi}_{ij} \right\rangle_\Omega = 0, \quad
  \left\langle \tilde{\varphi}_{ij} \tilde{\varphi}_{kl}
  \right\rangle_\Omega =
  \frac{\sigma^2}{15}
  \left(
      \delta_{ik} \delta_{jl} +  \delta_{il} \delta_{jk}
      - \frac{2}{3} \delta_{ij} \delta_{kl}
    \right) J_2,
\end{equation}
the angular averages in the integrand of Eq.~(\ref{eq:a-19}) for
$n=1,2$ are given by
\begin{align}
  \label{eq:a-21a}
  \left\langle
  \mathcal{D}(\bm{k}) \mathcal{P}
  \right\rangle_\Omega
  &= \frac{W(kR)}{\sigma} \frac{\partial}{\partial J_1} 
    \mathcal{P}
    = - \frac{W(kR)}{\sigma} J_1 \mathcal{P},
\\
  \label{eq:a-21b}
  \left\langle
  \mathcal{D}(\bm{k}_1)\mathcal{D}(\bm{k}_2)
  \mathcal{P}
  \right\rangle_\Omega
  &=
  \frac{W(k_1R)W(k_2R)}{\sigma^2}
  \left\{
  \frac{\partial^2}{\partial {J_1}^2}
  +
  \left[
  3 (\hat{\bm{k}}_1\cdot\hat{\bm{k}}_2)^2 - 1
  \right]
  \left( 1 + \frac{2}{5} J_2 \frac{\partial}{\partial J_2} \right)
  \frac{\partial}{\partial J_2}
  \right\} \mathcal{P}
  \nonumber\\
  &=
  \frac{W(k_1R)W(k_2R)}{\sigma^2}
  \left\{
  {J_1}^2 - 1
  +
  \left[
  3 (\hat{\bm{k}}_1\cdot\hat{\bm{k}}_2)^2 - 1
  \right]
  \frac{5}{2} \left( J_2 - 1\right)
  \right\} \mathcal{P}.
\end{align}
Substituting Eqs.~(\ref{eq:a-21a}) and (\ref{eq:a-21b}) into
Eq.~(\ref{eq:a-19}), the first- and second-order renormalized bias
functions are derived as
\begin{align}
  \label{eq:a-22a}
  c_1(\bm{k})
  &= b^\mathrm{L}_1 W(kR),
  \\              
  \label{eq:a-22b}
  c_2(\bm{k}_1,\bm{k}_2)
  &=
    \left\{
    b^\mathrm{L}_2 + 
    \left[
    3 (\hat{\bm{k}}_1\cdot\hat{\bm{k}}_2)^2 - 1
    \right] \omega^\mathrm{L}_1
    \right\}
    W(k_1R)W(k_2R),
\end{align}
where
\begin{equation}
  \label{eq:a-23}
  b^\mathrm{L}_1 \equiv
  \frac{1}{\sigma \bar{n}_\mathrm{PBH}}
  \left\langle J_1 n_\mathrm{PBH} \right\rangle, \quad
  b^\mathrm{L}_2 \equiv
  \frac{1}{\sigma^2 \bar{n}_\mathrm{PBH}}
  \left\langle \left({J_1}^2 - 1\right) n_\mathrm{PBH}\right\rangle,
  \quad
  \omega^\mathrm{L}_1 \equiv
  \frac{1}{\sigma^2 \bar{n}_\mathrm{PBH}}
  \left\langle \frac{5}{2}\left(J_2 - 1\right) n_\mathrm{PBH}\right\rangle,
\end{equation}
and $\langle\cdots\rangle = \int d^6\!y\cdots \mathcal{P}(\bm{y})$.
General definitions of $b^\mathrm{L}_n$ and $\omega^\mathrm{L}_l$ are
given by \cite{MD16}
\begin{equation}
  \label{eq:a-24}
  b^\mathrm{L}_n \equiv
  \frac{1}{\sigma^n \bar{n}_\mathrm{PBH}}
  \left\langle H_n\left(J_1\right) n_\mathrm{PBH}\right\rangle,
  \quad
  \omega_l \equiv
  \frac{(-1)^l}{\sigma^{2l} \bar{n}_\mathrm{PBH}}
  \left\langle
  L^{(3/2)}_l\left(\frac{5}{2} J_2\right) n_\mathrm{PBH}
  \right\rangle,
\end{equation}
where $H_n(x)$ is the Hermite polynomial and $L^{(\alpha)}_l(x)$ is
the generalized Laguerre polynomial.

The variables $J_1$ and $J_2$, which are defined
by Eq.~(\ref{eq:a-14}), are represented by eigenvalues of
$\varphi_{ij}$ as
\begin{equation}
  \label{eq:a-26}
  J_1(\alpha,\beta,\gamma) = \frac{\alpha+\beta+\gamma}{\sigma}, \quad
  J_2(\alpha,\beta,\gamma) =
   \frac{    \alpha^2 + \beta^2 + \gamma^2
    - \alpha\beta - \beta\gamma - \gamma\alpha}{\sigma^2}.
\end{equation}
Using the distribution function $w(\alpha,\beta,\gamma)$ of
Eq.~(\ref{eq:3-3}), Eq.~(\ref{eq:a-23}) reduces to
\begin{align}
  \label{eq:a-27a}
  b^\mathrm{L}_1
  &\equiv
  \frac{1}{\sigma\beta_0}
  \int_0^\infty d\alpha \int_{-\infty}^\alpha d\beta
  \int_{-\infty}^\beta d\gamma
  J_1(\alpha,\beta,\gamma)
     \Theta\left[1-h(\alpha,\beta,\gamma)\right]
    w(\alpha,\beta,\gamma),
  \\
  \label{eq:a-27b}
  b^\mathrm{L}_2
  &\equiv
  \frac{1}{\sigma^2\beta_0}
  \int_0^\infty d\alpha \int_{-\infty}^\alpha d\beta
  \int_{-\infty}^\beta d\gamma
  \left[J_1(\alpha,\beta,\gamma)\right]^2
  \Theta\left[1-h(\alpha,\beta,\gamma)\right]
  w(\alpha,\beta,\gamma)
   - \frac{1}{\sigma^2},
  \\
  \label{eq:a-27c}
  \omega^\mathrm{L}_1
  &\equiv
  \frac{5}{2\sigma^2 \beta_0}
  \int_0^\infty d\alpha \int_{-\infty}^\alpha d\beta
  \int_{-\infty}^\beta d\gamma
  J_2(\alpha,\beta,\gamma)
  \Theta\left[1-h(\alpha,\beta,\gamma)\right]
    w(\alpha,\beta,\gamma)
    - \frac{5}{2\sigma^2}.
\end{align}

\section{\label{app:AnalyticRBF}
  Analytic estimates of renormalized bias functions
}

In this Appendix, the three-dimensional integrations of the previous
Appendix are reduced to two-dimensional integrals. After that,
analytic estimates for the coefficients $b^\mathrm{L}_1$,
$b^\mathrm{L}_2$, and $\omega^\mathrm{L}_1$ of
Eqs.~(\ref{eq:a-27a})--(\ref{eq:a-27c}) are derived in a limit of
$\sigma \ll 1$. Analytic estimates with the effects of angular
momentum are also presented.

\subsection{Without the effects of angular momentum}

According to Ref.~\cite{Har16},
it is useful to define new variables,
\begin{equation}
  \label{eq:b-1}
  t = \frac{2}{3} \frac{\alpha + \beta + \gamma}{\alpha - \gamma}, \quad
  u = \frac{1}{2} \frac{\alpha - 2\beta + \gamma}{\alpha - \gamma}, \quad
  z = \frac{\alpha - \gamma}{2}.
\end{equation}
The domain of integration, $\alpha > \beta > \gamma$, $\alpha > 0$ in
Eqs.~(\ref{eq:a-27a})--(\ref{eq:a-27c}) corresponds to $-1/2 < u < 1/2$,
$t > -(1 + 2u/3)$ and $z>0$. The condition $h(\alpha,\beta,\gamma) <
1$ is equivalent to
\begin{equation}
  \label{eq:b-2}
  z > z_*(t,u) \equiv
  \frac{4}{\pi}\left(t + \frac{2}{3}u + 1\right)^{-2}
  E\left[\sqrt{1 - \left(u + \frac{1}{2}\right)^2}\right].
\end{equation}
With new variables, we have
\begin{equation}
  \label{eq:b-3}
  w(\alpha,\beta,\gamma) d\alpha\,d\beta\,d\gamma  =
  \frac{3^3\cdot 5^{5/2}}{\pi \sigma^6}
  (1+2u)(1-2u) z^5 e^{-A(t,u)z^2} dt\,du\,dz,
\end{equation}
where
\begin{equation}
  \label{eq:b-4}
  A(t,u) \equiv
  \frac{1}{\sigma^2}
  \left(\frac{9}{2} t^2 + 10 u^2 + \frac{15}{2}\right).
\end{equation}
Defining the integrals,
\begin{equation}
  \label{eq:b-5}
  I_{nml} \equiv
  \int_{-1/2}^{1/2}du (1-2u)(1+2u) u^n
  \int_{-(1+2u/3)}^\infty dt\,(3t)^m
  \int_{z_*(t,u)}^\infty dz\, z^{l+5}
  e^{-A(t,u)z^2},
\end{equation}
Eqs.~(\ref{eq:3-5}) and (\ref{eq:a-27a})--(\ref{eq:a-27c}) reduce to
\begin{equation}
  \label{eq:b-6}
  \beta_0
  = \frac{3^3\cdot 5^{5/2}}{\pi\,\sigma^6} I_{000}, \qquad
  b^\mathrm{L}_1
  = \frac{I_{011}}{\sigma^2 I_{000}}, \qquad
  b^\mathrm{L}_2
  = \frac{1}{\sigma^2}
    \left(\frac{I_{022}}{\sigma^2 I_{000}} - 1 \right), \qquad
  \omega^\mathrm{L}_1
  = \frac{5}{2\sigma^2}
  \left(\frac{4I_{202} + 3 I_{002}}{\sigma^2 I_{000}} - 1\right).
\end{equation}
These expressions are the exact transformation of the original
integrals, Eqs.~(\ref{eq:3-5}) and (\ref{eq:a-27a})--(\ref{eq:a-27c}),
without any approximation. The last integral over $z$ in
Eq.~(\ref{eq:b-5}) can be analytically evaluated as
\begin{equation}
  \label{eq:b-7}
  \int_{z_*(t,u)}^\infty dz\, z^{l+5} e^{-A z^2} =
  A^{-(l+6)/2} \int_{\sqrt{A} z_*}^\infty dr\,r^{l+5} e^{-r^2} =
  \frac{1}{2} A^{-(l+6)/2} \Gamma\left(\frac{l}{2} + 3, A {z_*}^2\right),
\end{equation}
where $\Gamma(x)$ is the gamma function. 

Next, we derive the analytic estimates of the integrals of
Eq.~(\ref{eq:b-5}) in a limit $\sigma \ll 1$. In this limit, the
integral over $t$ is dominantly contributed by a region
$t \gtrsim \sigma^{-1}$, since the contribution from
$t \lesssim \sigma^{-1}$ is exponentially suppressed \cite{Har16}.
Therefore, the dominant contribution comes from
$t\gtrsim \sigma^{-1} \gg 1$, and in this region, we have
$A\simeq 9t^2/(2\sigma^2)$, $z_*\simeq 4E/(\pi t^2)$, and 
$A {z_*}^2 \simeq 72 E^2/(\pi^2 \sigma^2 t^2)$. The lower limit
$-(1+2u/3)$ of the integral over $t$ can be replaced by $0$ because
$t \gg 1$. Introducing variables $s = 6\sqrt{2}E/(\pi\sigma t)$ (with
$u$ fixed) and $r = \sqrt{A}z$ (with $t,u$ fixed), the last two
integrals in Eq.~(\ref{eq:b-5}) are approximately given by
\begin{equation}
  \label{eq:b-8}
  \int_{-(1+2u/3)}^\infty dt\,(3t)^m
  \int_{z_*(t,u)}^\infty dz\, z^{l+5}
  e^{-A(t,u)z^2}
  \simeq
  \frac{(2\sigma^2)^{l/2+3}}{3^{l-m+6}}
  \left(\frac{\pi\sigma}{6\sqrt{2}E}\right)^{l-m+5}
  \int_0^\infty ds\,s^{l-m+4} \int_s^\infty dr\,r^{l+5} e^{-r^2},
\end{equation}
where the argument of the elliptic integral $E$ is the same as in
Eq.~(\ref{eq:b-2}). Using a partial
integration, the above integral reduces to an analytic form with a
gamma function. As a result, Eq.~(\ref{eq:b-5}) in the case of $\sigma
\ll 1$ is given by
\begin{equation}
  \label{eq:b-9}
  I_{nml} \simeq
  \frac{\Gamma(l-m/2+11/2)}{l-m+5}
  2^{-l+3m/2-9/2}\cdot 3^{-2l+2m-12}
  \pi^{l-m+5} \sigma^{2l-m+11}
  \frac{3}{2}
  \int_{-1/2}^{1/2}du (1-2u)(1+2u) u^n E^{-l+m-5}.
\end{equation}

Substituting Eq.~(\ref{eq:b-9}) into Eq.~(\ref{eq:b-6}), we have
\begin{equation}
  \label{eq:b-10}
  \beta_0
  \simeq \frac{5^3\cdot 7 \pi^{9/2}}{2^9\cdot 3^6\sqrt{10}}
    \bar{E}^{-5} \sigma^5 \simeq 0.05556\sigma^5,
\end{equation}
where
\begin{equation}
  \label{eq:b-11}
  \bar{E}^{-5} \equiv
  \frac{3}{2}
  \int_{-1/2}^{1/2}du (1-2u)(1+2u) E^{-5},
\end{equation}
which is already known in Ref.~\cite{Har16}. Substituting
Eq.~(\ref{eq:b-9}) into Eq.~(\ref{eq:b-6}), we have
\begin{equation}
  \label{eq:b-12}
  b^\mathrm{L}_1
  \simeq
    \frac{2^8\sqrt{2}}{3^2\cdot 7\sqrt{\pi}\sigma}
    \simeq \frac{3.242}{\sigma}, \qquad
  b^\mathrm{L}_2
  \simeq \frac{10}{\sigma^2}, \qquad
  \omega^\mathrm{L}_1
  \simeq - \frac{5}{2\sigma^2},
\end{equation}
where only dominant terms of $\sigma \ll 1$ are retained.

\begin{figure}[t]
  \begin{minipage}[t]{0.45\linewidth}
    \centering
    \includegraphics[width=19pc]{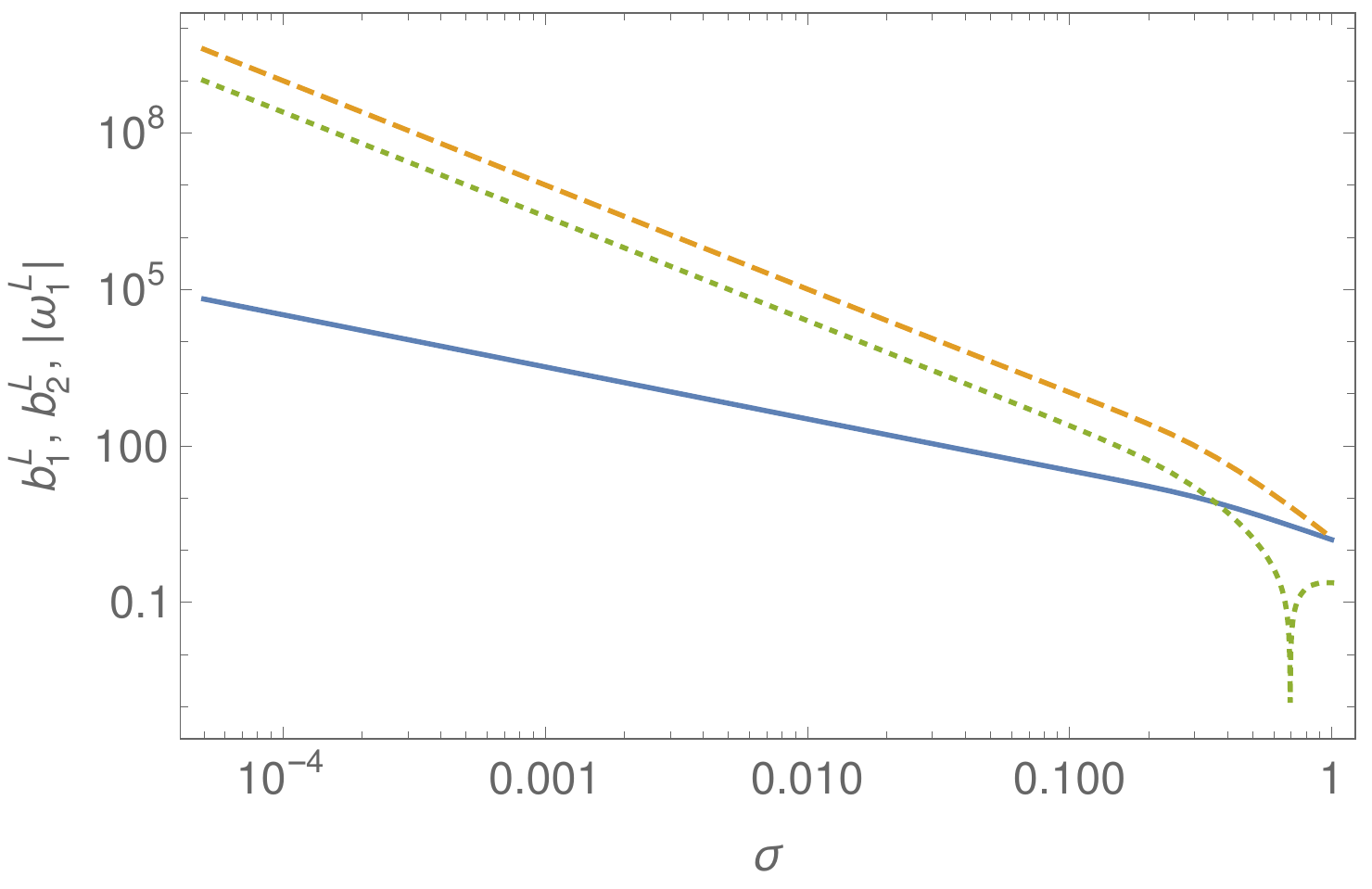}
    \raggedleft
    \caption{\label{fig:coeffwo} Bias coefficients $b^\mathrm{L}_1$
      (blue solid line), $b^\mathrm{L}_2$ (orange dashed line), and
      $\omega^\mathrm{L}_1$ (green dotted line) in the model of
      Ref.~\cite{Har16}. The parameter $\omega^\mathrm{L}_1$ is
      negative except for $\sigma \gtrsim 0.65$. }
  \end{minipage}
  \hspace{1pc}
  \begin{minipage}[t]{0.45\linewidth}
    \centering
    \includegraphics[width=19pc]{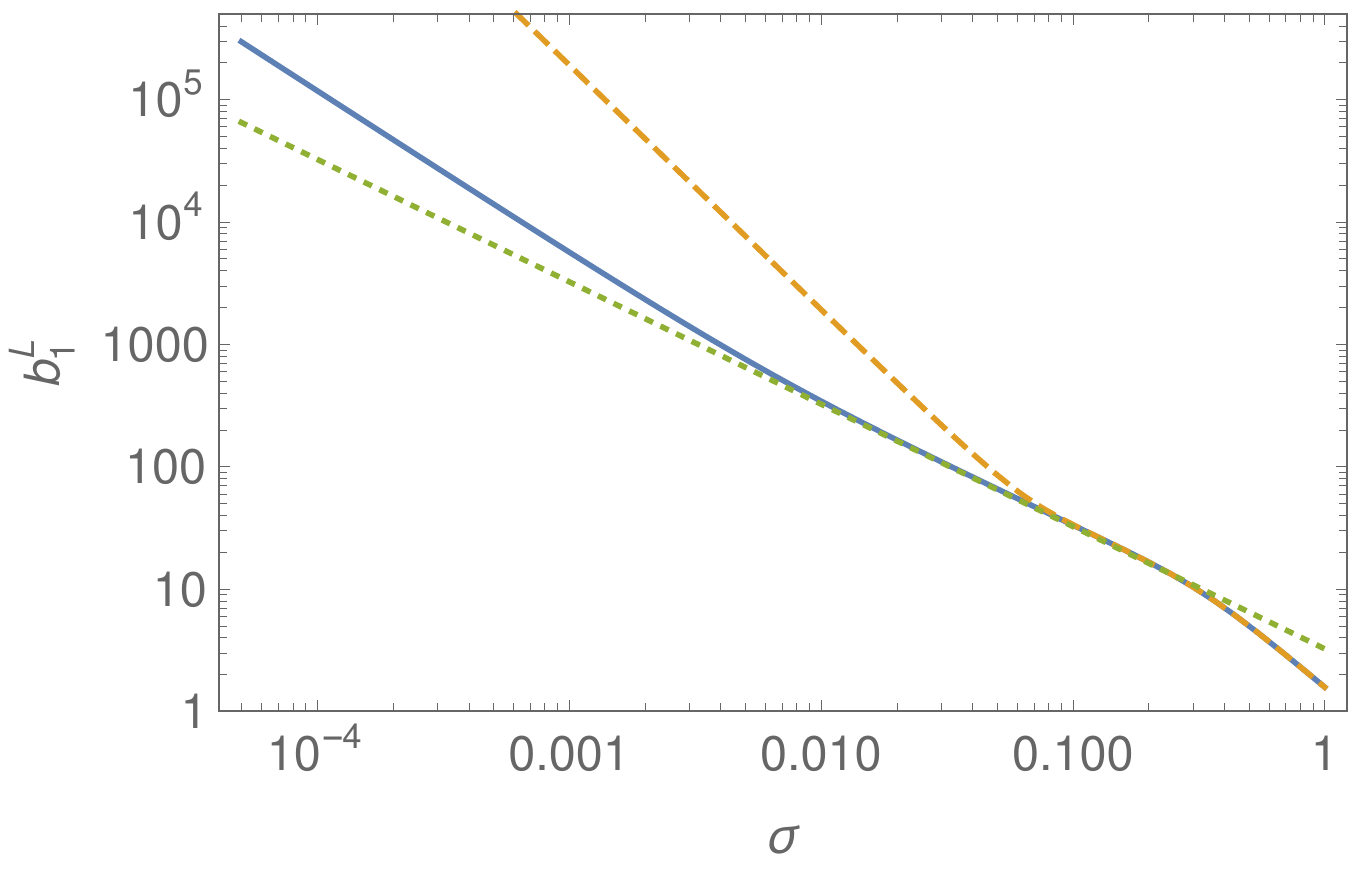}
    \raggedleft
    \caption{\label{fig:b1wa} The bias coefficient $b^\mathrm{L}_1$
      for second-order case (blue solid line), first-order case
      (orange dashed line) and the asymptotic formula of
      $\sigma \ll 1$ without the effects of angular momentum (green
      dotted line). }
  \end{minipage}
\end{figure}

In Fig.~\ref{fig:coeffwo}, the bias coefficients $b^\mathrm{L}_1$,
$b^\mathrm{L}_2$, and $\omega^\mathrm{L}_1$ as functions of $\sigma$
are plotted. Numerical integrations are performed by
Eq.~(\ref{eq:b-6}) with Eqs.~(\ref{eq:b-5}) and (\ref{eq:b-7}). These
coefficients for $\sigma \lesssim 0.1$ have power-law shapes, which
are well described by Eq.~(\ref{eq:b-12}).

\begin{figure}[t]
  \begin{minipage}[t]{0.45\linewidth}
    \centering
    \includegraphics[width=19pc]{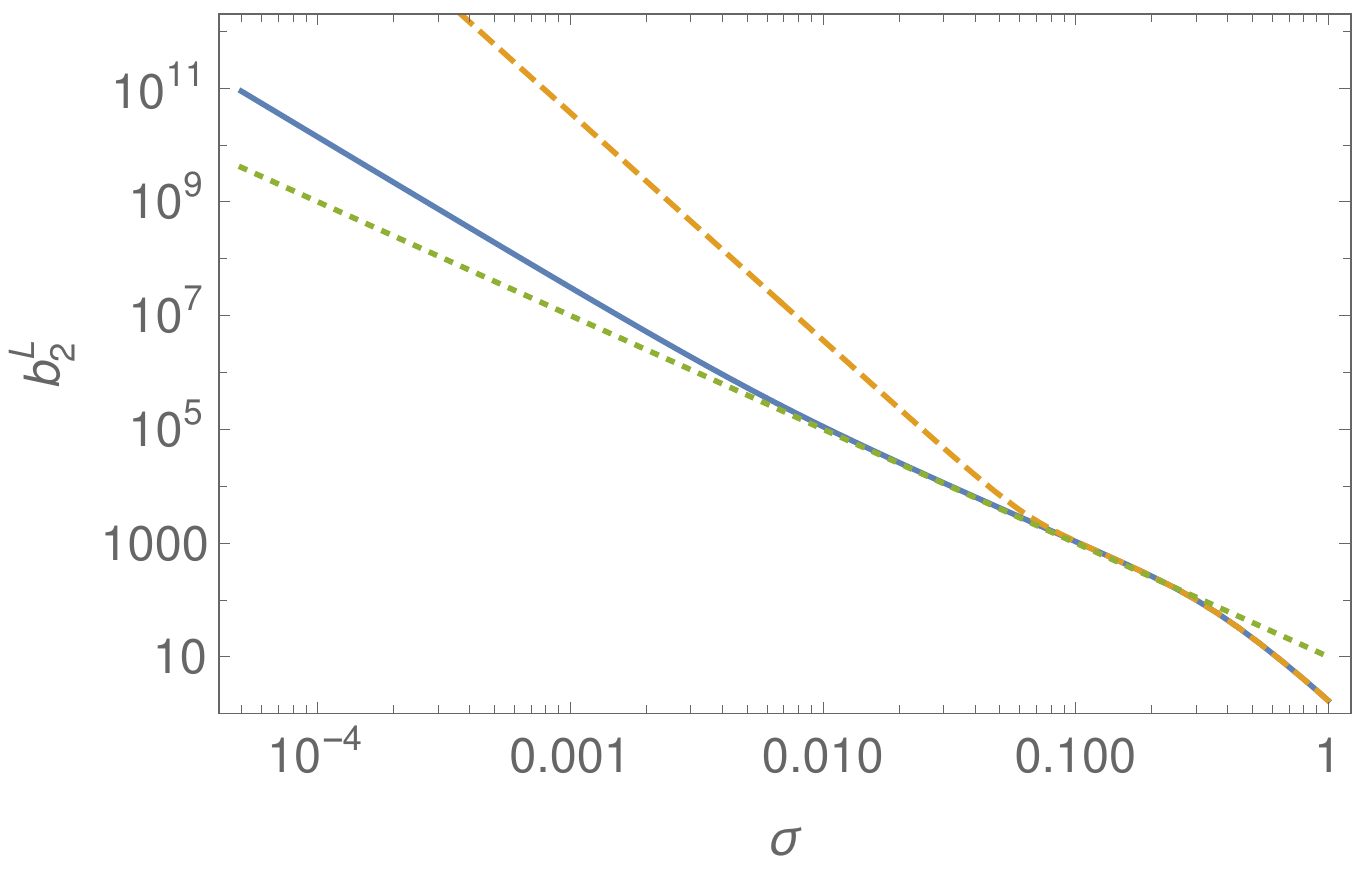}
    \caption{\label{fig:b2wa} Same as in Fig.~\ref{fig:b1wa} but for
      the bias coefficient $b^\mathrm{L}_2$. }
  \end{minipage}
  \hspace{1pc}
  \begin{minipage}[t]{0.45\linewidth}
    \centering
    \includegraphics[width=19pc]{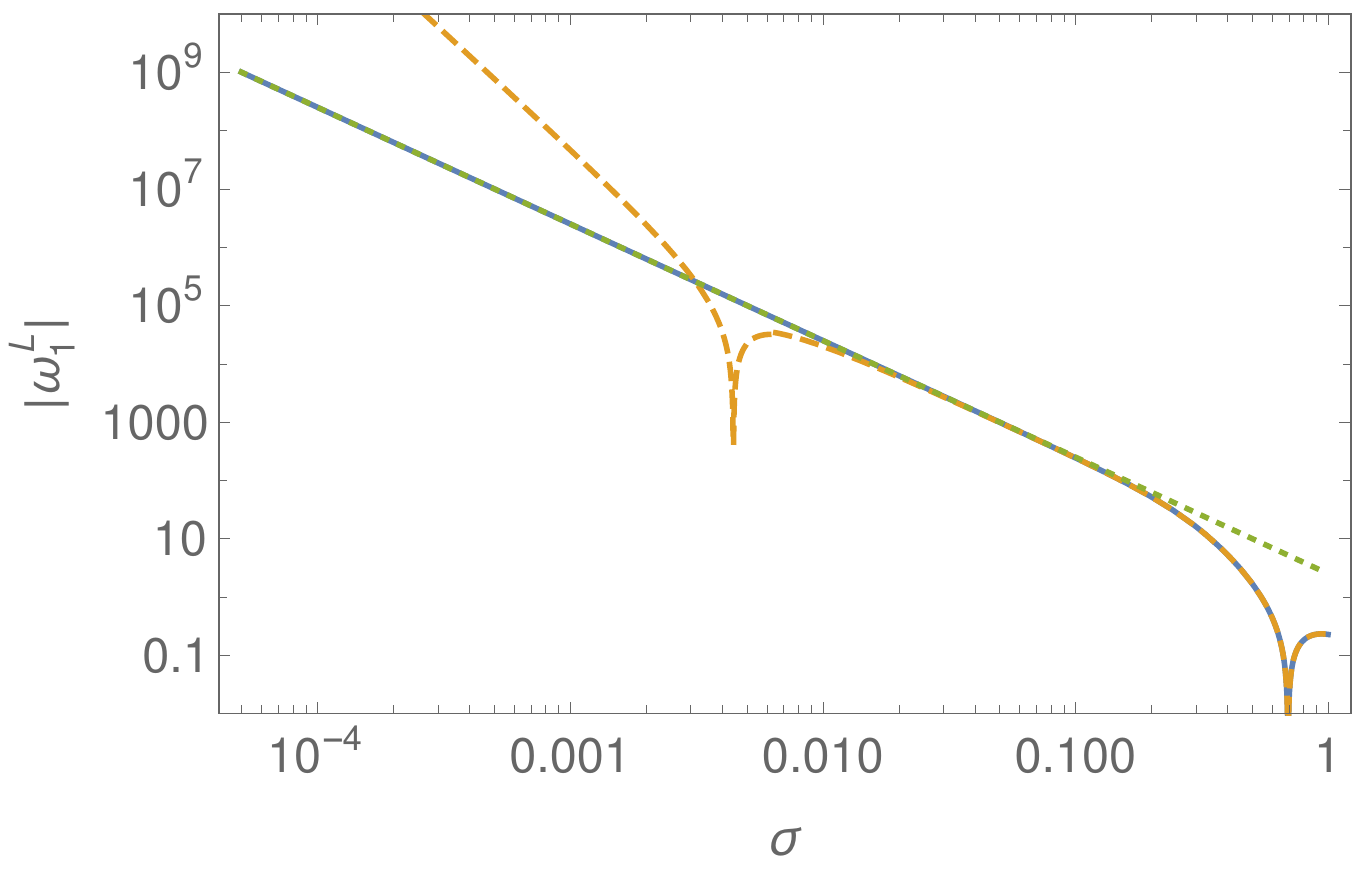}
    \caption{\label{fig:omwa} Same as in Fig.~\ref{fig:b1wa} but for
      the bias coefficient $\omega^\mathrm{L}_1$. }
  \end{minipage}
\end{figure}

\subsection{Effects of angular momentum}

We briefly give the analytic estimates including the effects of
angular momentum. In this case, the integrands of Eqs.~(\ref{eq:3-5})
and (\ref{eq:a-27a})--(\ref{eq:a-27c}) are multiplied by a factor,
$\Theta(\alpha + \beta + \gamma - \delta_\mathrm{th})$. The function
$z_*(t,u)$ in Eqs.~(\ref{eq:b-5}) and (\ref{eq:b-7}) is substituted by
$z_*(t,u) \rightarrow z_0(t,u) \equiv \mathrm{max}[z_*(t,u),
z_\mathrm{th}(t)]$, where
$z_\mathrm{th} \equiv \delta_\mathrm{th}/(3t)$. Extending the
derivation of the previous subsection, and using the similar
considerations of Ref.~\cite{Har17}, one can derive
\begin{equation}
  \label{eq:b-21}
  I_{nml} \simeq
  \frac{2^{-2l+2m-9}\cdot 3^{-2l+2m-12}\pi^{l-m+5}}{l-m+5}\,
    {\delta_\mathrm{th}}^{2l-m+9} \sigma^2e^{-\delta_\mathrm{th}^2/2\sigma^2}
  \frac{3}{2}
  \int_{-1/2}^{1/2}du (1-2u)(1+2u) u^n E^{-l+m-5},
\end{equation}
for $\sigma \ll\delta_\mathrm{th} \ll 1$.  Substituting the above equation into
Eq.~(\ref{eq:b-6}), we finally have
\begin{align}
  \label{eq:b-22a}
  &\beta_0 \simeq \frac{5^{3/2}\pi^4}{2^9\cdot 3^9}
  \bar{E}^{-5} \frac{{\delta_\mathrm{th}}^9}{\sigma^4}
  \exp\left(-\frac{{\delta_\mathrm{th}}^2}{2\sigma^2}\right) \simeq
  4.691 \times 10^{-5}
   \frac{{\delta_\mathrm{th}}^9}{\sigma^4}
  \exp\left(-\frac{{\delta_\mathrm{th}}^2}{2\sigma^2}\right),\\
  \label{eq:b-22b}
  & b^\mathrm{L}_1 \simeq \frac{\delta_\mathrm{th}}{\sigma^2},\qquad
  b^\mathrm{L}_2 \simeq \frac{{\delta_\mathrm{th}}^2}{\sigma^4}, \qquad
  \omega^\mathrm{L}_1 \simeq 0.03400
  \frac{{\delta_\mathrm{th}}^4}{\sigma^4}  - \frac{5}{2\sigma^2}.
\end{align}
The Eq.~(\ref{eq:b-22a}) agrees with a result of Ref.~\cite{Har17}.

In Figs.~\ref{fig:b1wa}, \ref{fig:b2wa} and \ref{fig:omwa}, the
results of numerical integrations for bias coefficients
$b^\mathrm{L}_1$, $b^\mathrm{L}_2$, and $\omega^\mathrm{L}_1$ are
plotted, respectively. It is difficult to accurately evaluate the
numerical integrations of the first-order case for
$\sigma \lesssim 0.007$, where the production probability is
significantly suppressed, and the lines are artificially connected to
asymptotic formulas, Eqs.~(\ref{eq:b-22b}). In the second-order case,
the effects of angular momentum are small for $\sigma \gtrsim 0.01$.

\section{More precise discussion on the observational constraints
  \label{app:constraints}
}

\begin{figure}[t] 
\begin{center}
\includegraphics[width=0.5\columnwidth]{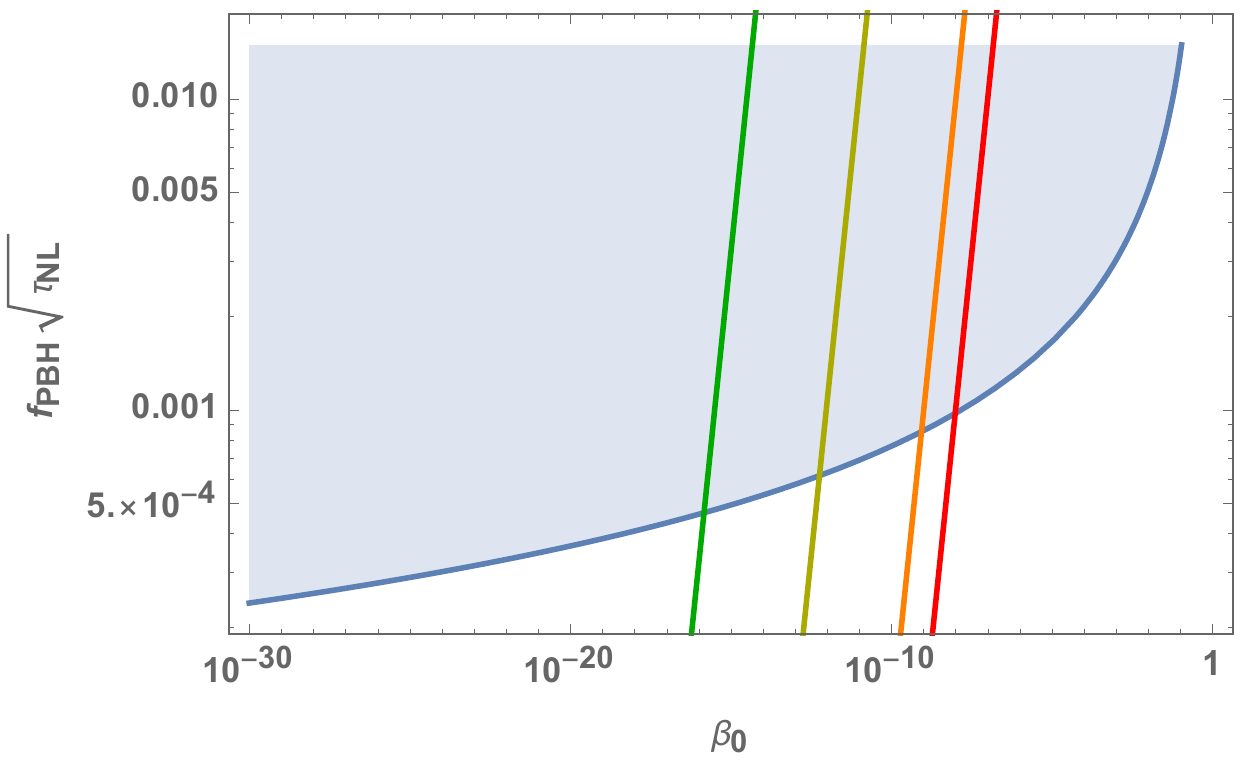}
\caption{ Upper bound on $f_\text{PBH} \sqrt{\tau_\text{NL}}$ as a
  function of $\beta_0$ in the case of PBH production in a RD epoch
  (blue curve). Four examples of $f_\text{PBH}\sqrt{\tau_\text{NL}}$
  are plotted as functions of $\beta_0$ with the same color coding as
  those in Fig.~\ref{fig:vsShotNoise}:
  $M_\text{PBH}/M_\odot = 10^{-12}$ (green; DM), $10^{-5}$ (dark
  yellow; OGLE), $10^{1.5}$ (orange), and $10^{3.5}$ (red; SMBH). }
\label{fig:max_fSqrtTau_RD}
\end{center}
\end{figure}

In the main text, we have just used $C_2 \gtrsim 20$ and $C_2 \geq 5$
to place constraints in the RD case and the MD case, respectively.
More precisely, $C_2$ depends nonlinearly on the production
probability $\beta_0$, while $f_\text{PBH}$ itself depends linearly on
$\beta_0$. 
We have already seen in Fig.~\ref{fig:max_fSqrtTau_MD} that the upper
bound on $f_{\text{PBH}} \sqrt{\tau_\text{NL}}$ has a nontrivial shape
as a function of $\beta_0$ in the case of PBH production in the MD
epoch. The counterpart in the RD epoch is shown in
Fig.~\ref{fig:max_fSqrtTau_RD}.

Thus, the constraint~\eqref{eq:f_upper-bound} is a nonlinear
constraint on $\beta_0$ depending on $\tau_\text{NL}$ and the
temperature at which the scale corresponding to the PBH mass
$M_\text{PBH}$ enters the horizon (RD case), or the reheating
temperature $T_\text{R}$ (MD case). We can obtain upper bounds on
$f_\text{PBH}$ as a function of $M_\text{PBH}$ (RD) or $T_\text{R}$
(MD) for fixed $\tau_\text{NL}$ and as a function of $\tau_\text{NL}$
for fixed $M_\text{PBH}$ (RD) or $T_\text{R}$ (MD).

\begin{figure}[t]
  \begin{minipage}[t]{0.45\linewidth}
    \centering
    \includegraphics[width=19pc]{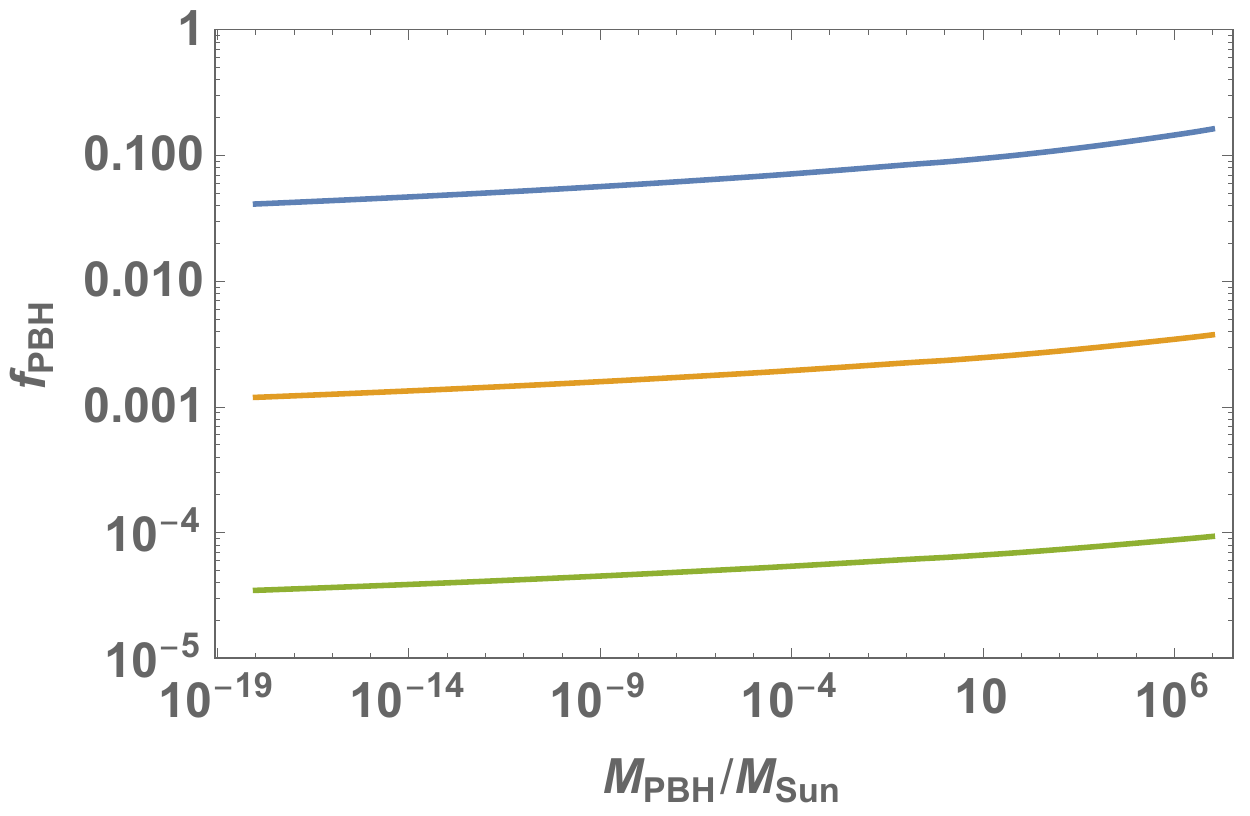}
    \caption{ Upper bounds on $f_\text{PBH}$ as a function of
      $M_\text{PBH}$ for $\tau_\text{NL}= 10^{-4}$ (blue line),
      $10^{-1}$ (orange line), and $10^2$ (green line) in the case of
      PBH production in a RD epoch. }
    \label{fig:max_f_M_RD}
  \end{minipage}
  \hspace{1pc}
  \begin{minipage}[t]{0.45\linewidth}
    \centering
    \includegraphics[width=19pc]{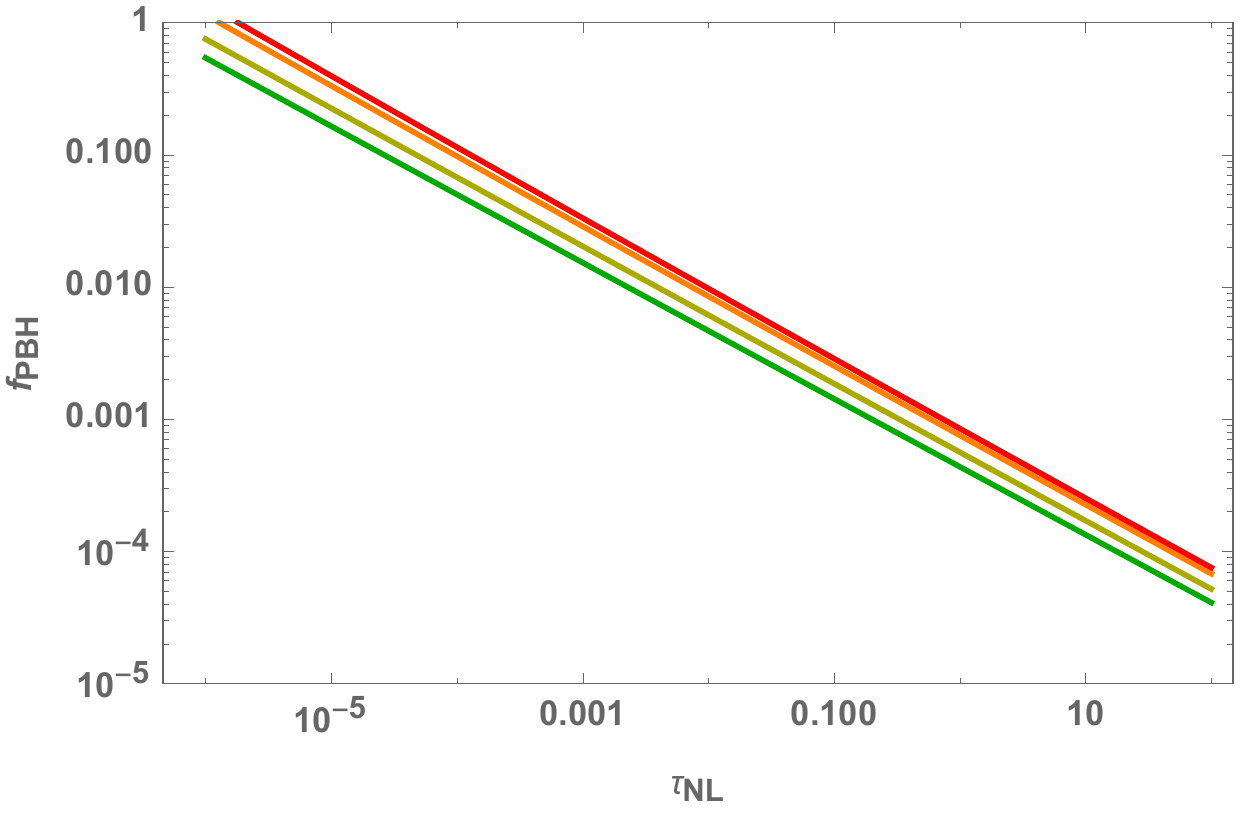}
    \caption{ Upper bounds on $f_\text{PBH}$ as a function of
      $\tau_\text{NL}$ for $M_\text{PBH}/M_\odot= 10^{-12}$ (green;
      DM), $10^{-5}$ (dark yellow; OGLE), $10^{1.5}$ (orange; LIGO/Virgo),
      and $10^{3.5}$ (red; SMBH) in the case of PBH production in a RD
      epoch. }
    \label{fig:max_f_tau_RD}
  \end{minipage}
\end{figure}

In the case of PBH production in a RD epoch, an essentially same upper
bound on $f_\text{PBH} (M_\text{PBH})$ for fixed $f_\text{NL}$
(instead of $\tau_\text{NL}$) is given in Fig.~2 of Ref.~\cite{TY15}.
For completeness and with updated Planck data, we show a similar upper
bound on $f_\text{PBH}(M_\text{PBH})$ for some choices of
$\tau_\text{NL}$ in Fig.~\ref{fig:max_f_M_RD}. The fact that the mass
dependence is weak is also shown in Fig~\ref{fig:max_f_tau_RD}, in
which the upper bound is shown as a function of $\tau_\text{NL}$ for
fixed masses. To obtain these figures, we solved the relation between
$C_2$ and $\beta_0$ using Eqs.~\eqref{eq:2-28} and \eqref{eq:2-29-1}.

\begin{figure}[t]
  \begin{minipage}[t]{0.45\linewidth}
    \centering
    \includegraphics[width=19pc]{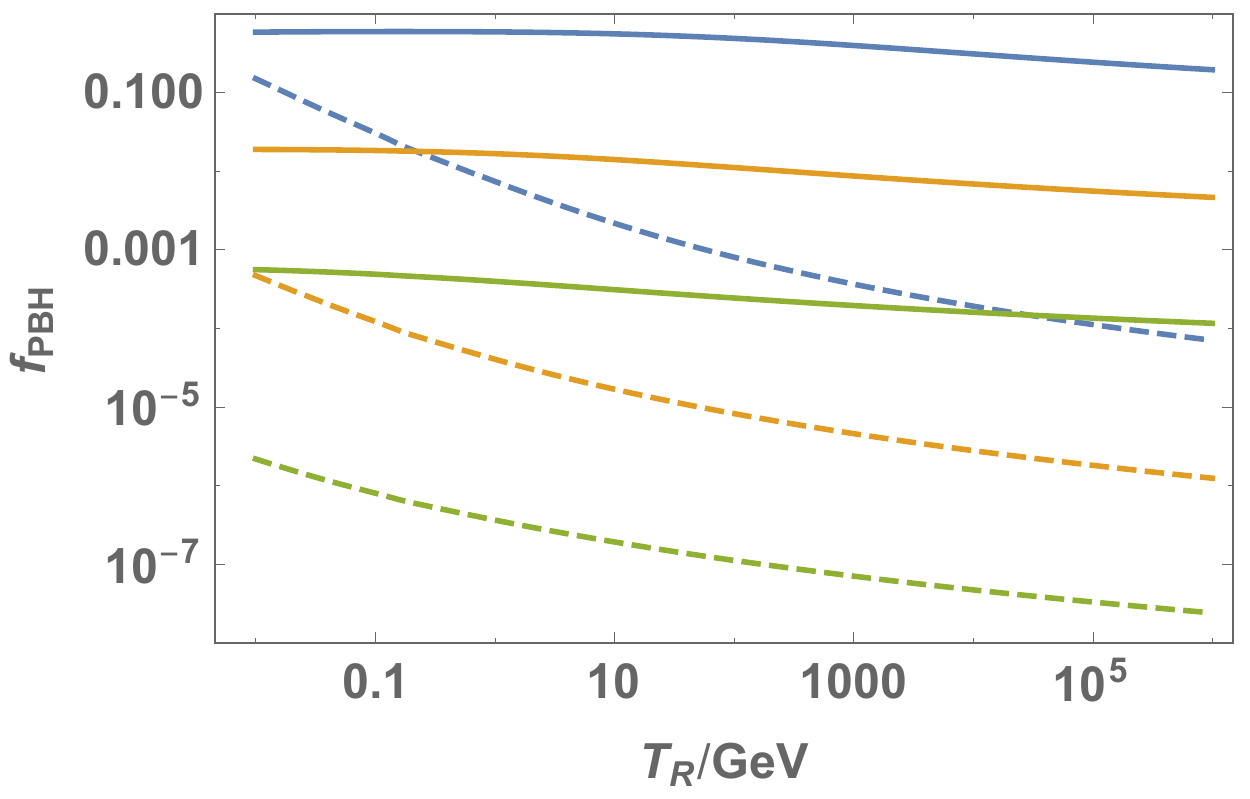}
    \caption{ Upper bounds on $f_\text{PBH}$ as a function of
      $T_\text{R}$ for $\tau_\text{NL}= 10^{-4}$ (blue lines),
      $10^{-1}$ (orange lines), and $10^2$ (green lines) in the case
      of PBH production in a MD epoch. The solid (dashed) lines are
      based on the second-order (first-order) production mechanism. }
    \label{fig:max_f_TR_MD}
  \end{minipage}
  \hspace{1pc}
  \begin{minipage}[t]{0.45\linewidth}
    \centering
    \includegraphics[width=19pc]{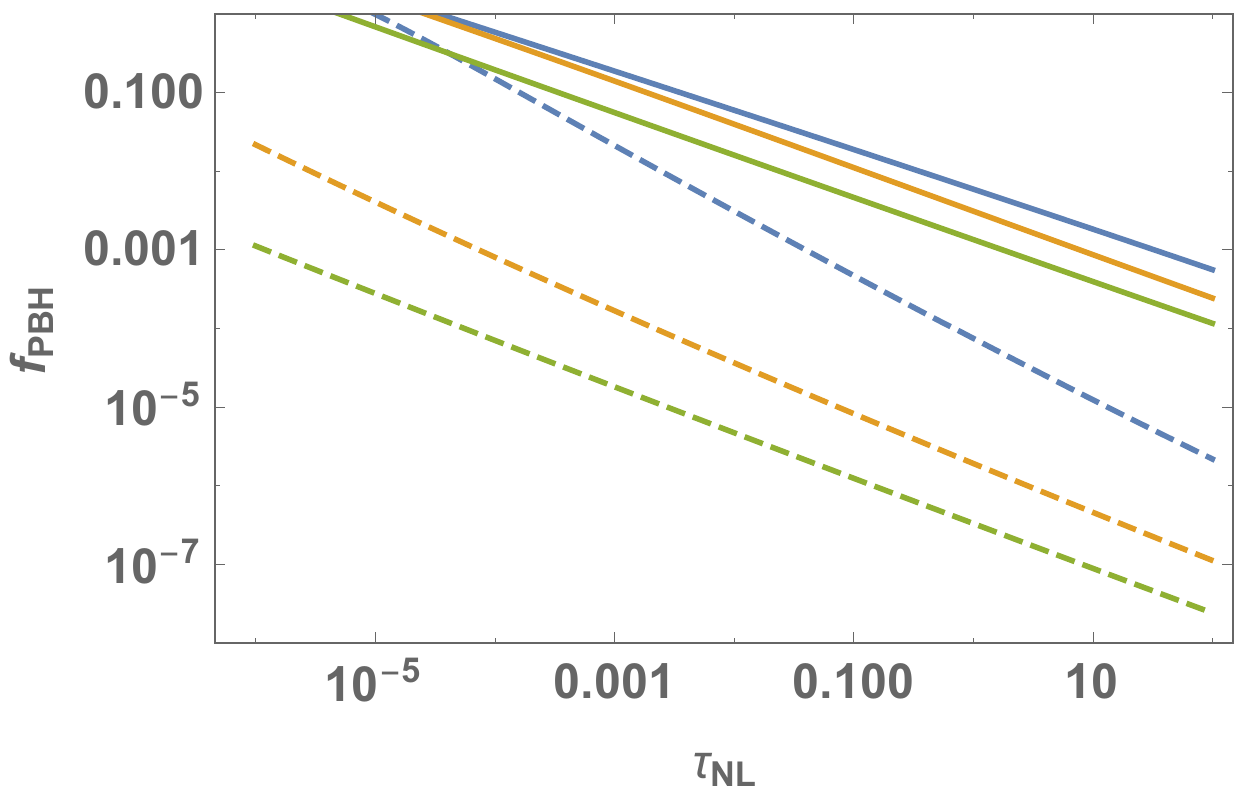}
    \caption{ Upper bounds on $f_\text{PBH}$ as a function of
      $\tau_\text{NL}$ for $T_\text{R} = 10^{-2}$~GeV (blue lines),
      $10^{2}$~GeV (orange lines), and $10^6$~GeV (green lines) in the
      case of PBH production in a MD epoch. The solid (dashed) lines are
      based on the second-order (first-order) production mechanism. }
    \label{fig:max_f_tau_MD}
  \end{minipage}
\end{figure}

In the case of PBH production in a MD epoch, the isocurvature constraint
on $f_\text{PBH}$ is given in terms of $T_\text{R}$ instead of
$M_\text{PBH}$. This is shown in Fig.~\ref{fig:max_f_TR_MD}. The
constraint as a function of $\tau_\text{NL}$ for fixed $T_\text{R}$ is
shown in Fig.~\ref{fig:max_f_tau_MD}. To obtain these figures, the
relations between $C_2$ and $\beta_0$ were numerically solved using
results in the main text as shown in Fig.~\ref{fig:C2vsBeta}.

\newpage

\renewcommand{\apj}{Astrophys.~J. }
\newcommand{\aap}{Astron.~Astrophys. }
\newcommand{\aj}{Astron.~J. }
\newcommand{\apjl}{Astrophys.~J.~Lett. }
\newcommand{\apjs}{Astrophys.~J.~Suppl.~Ser. }
\newcommand{\apss}{Astrophys.~Space Sci. }
\newcommand{\cqg}{Class.~Quant.~Grav. }

\newcommand{\jcap}{J.~Cosmol.~Astropart.~Phys. }
\newcommand{\mnras}{Mon.~Not.~R.~Astron.~Soc. }
\newcommand{\mpla}{Mod.~Phys.~Lett.~A }
\newcommand{\pasj}{Publ.~Astron.~Soc.~Japan }
\newcommand{\physrep}{Phys.~Rep. }
\newcommand{\ptp}{Progr.~Theor.~Phys. }
\newcommand{\ptep}{Prog.~Theor.~Exp.~Phys. }
\newcommand{\jetp}{JETP }


\twocolumngrid

\end{document}